\documentclass[aps, prd,linenumbers, twocolumn,superscriptaddress,preprintnumbers]{mnras}

\usepackage{natbib}
\include{notations}

\usepackage{amsmath,amssymb,latexsym,times}

\usepackage{numdef}
\usepackage{slashed}
\usepackage{graphicx}
\usepackage{grffile}
\usepackage[normalem]{ulem}
\usepackage{dcolumn}
\usepackage[dvipsnames]{xcolor}
\usepackage{booktabs}
\usepackage{amssymb}
\usepackage{slashed}
\usepackage{todonotes}
\usepackage{comment}
\usepackage{footnote}
\usepackage{newtxtext,newtxmath}

\usepackage[T1]{fontenc}
\usepackage{graphicx}	
\usepackage{amsmath}	
\usepackage{amssymb}	
\usepackage{slashed}
\usepackage[normalem]{ulem}
\usepackage{xspace}
\definecolor{aiiro}{HTML}{105779}
\usepackage{lineno}
\usepackage{hyperref}
\usepackage{threeparttable,booktabs}
\hypersetup{
     colorlinks=true,
     linkcolor=aiiro,
     citecolor = aiiro,      
     urlcolor=aiiro,
     }

\def\Cref#1{Chapter~\ref{#1}\xspace}

\graphicspath{{./}{figures/}}

\newcommand{\nk}{$\langle \delta_g \kappa_{\rm CMB}\rangle $}
\newcommand{\gk}{$\langle \gamma \kappa_{\rm CMB}\rangle$}
\newcommand{\fivetwo}{$5\! \times\! 2 {\rm pt}$}
\newcommand{\threetwo}{$3\! \times\! 2 {\rm pt}$}
\newcommand{\nkgk}{$\langle \delta_g \kappa_{\rm CMB}\rangle + \langle \gamma \kappa_{\rm CMB}\rangle$}

\newcommand{\planck}{{\it Planck}}

\newcommand{\sixtwo}{6\! \times\! 2 {\rm pt}\ }

\begin{document}

\title[\selectfont Galaxies$\times$CMB lensing: from Stage-III to Stage-IV]{{Transitioning from Stage-III to Stage-IV:\\ Cosmology from galaxy$\times$CMB lensing and shear$\times$CMB lensing}}

\date{\today}

\author[Z.~Zhang, C.~Chang et al.]{{
Zhuoqi (Jackie) Zhang$^{\star,1}$, 
Chihway Chang$^{\dagger,1,2}$, 
Patricia Larsen$^{3}$, 
Lucas F. Secco$^{2}$,
Joe Zuntz$^{4}$} 
\newauthor
{(The LSST Dark Energy Science Collaboration)}
\\ \\
$^{1}$ Department of Astronomy and Astrophysics, University of Chicago, Chicago, IL 60637, USA\\
$^{2}$ Kavli Institute for Cosmological Physics, University of Chicago, Chicago, IL 60637, USA\\
$^{3}$ Argonne National Laboratory, 9700 South Cass Avenue, Lemont, IL 60439, USA\\
$^{4}$ Institute for Astronomy, University of Edinburgh, Edinburgh EH9 3HJ, UK\\ \\
$^{\star}$ E-mail: zhuoqizhang@uchicago.edu\\
$^{\dagger}$ E-mail: chihway@kicp.uchicago.edu \\
}

\maketitle

\begin{abstract}
We examine the cosmological constraining power from two cross-correlation probes between galaxy and CMB surveys: the cross-correlation of lens galaxy density with CMB lensing convergence \nk{}, and source galaxy weak lensing shear with CMB lensing convergence \gk{}. These two cross-correlation probes provide an independent cross-check of other large-scale structure constraints and are insensitive to  galaxy-only or CMB-only systematic effects. In addition, when combined with other large-scale structure probes, the cross-correlations can break degeneracies in cosmological and nuisance parameters, improving both the precision and robustness of the analysis. In this work, we study how the constraining power of \nkgk{} changes from Stage-III (ongoing) to Stage-IV (future) surveys. Given the flexibility in selecting the \textit{lens} galaxy sample, we also explore systematically the impact on cosmological constraints when we vary the redshift range and magnitude limit of the lens galaxies using mock galaxy catalogs. We find that in our setup, the contribution to cosmological constraints from \nk{} and \gk{} are comparable in the Stage-III datasets; but in Stage-IV surveys, the noise in \nk{} becomes subdominant to cosmic variance, preventing \nk{} to further improve the constraints. This implies that to maximize the cosmological constraints from future \nkgk{} analyses, we should focus more on the requirements on \gk{} instead of \nk{}. Furthermore, the selection of the lens sample should be optimized in terms of our ability to characterize its redshift or galaxy bias instead of its number density.\\\hfill
\end{abstract}
\vspace{0.2in}

\section{Introduction}
\label{sec:introduction}
The cross correlation between pairs of cosmological tracers has become a powerful tool to constrain cosmology and probe new physics \citep{Schaan:2017,Schmittfull2018,Mishra-Sharma2018,y15x2,Yu2021,Krolewski2021,Fang2021,Chen2021}. As different tracers probe the same underlying large-scale structure while being sensitive to different cosmological and astrophysical/observational nuisance parameters, their cross-correlation can lead to more robust and precise constraints. One prominent example used in recent cosmological analyses is to combine the cross-correlation between galaxy positions $\delta_{g}$ and galaxy lensing shear $\gamma$ (a.k.a. galaxy-galaxy lensing) with the auto-correlation of galaxy positions (a.k.a. galaxy clustering) and galaxy lensing (a.k.a. cosmic shear). The techniques and modeling associated with combining these three sets of correlation functions, commonly referred to as the \threetwo{} probes, have matured significantly over the past 5 years with the application on state-of-the-art ``Stage-III\footnote{The Stage-III and Stage-IV classification was introduced in the Dark Energy Task Force report \citep{Albrecht2006}, where Stage-III refers to the ongoing dark energy experiments that started in the 2010s and Stage-IV refers to those that start in the 2020s.}'' galaxy surveys such as the Dark Energy Survey \citep[DES,][]{Flaugher2005} in \citet{y3-3x2ptkp}, and the Kilo-Degree Survey \citep[KiDS,][]{deJong2013} in \citet{Heymans2021}. A natural extension to the \threetwo{} analysis is to include CMB lensing in the same framework -- in particular, adding two-point functions that include CMB lensing convergence, $\kappa_{\rm CMB}$, as a third class of tracers. The combination of \threetwo{} with the galaxy density-CMB lensing and the galaxy shear-CMB lensing  cross correlations is sometimes referred to as the \fivetwo{} probes, and further adding CMB lensing auto-correlation becomes the $\sixtwo$ probes \citep[e.g.][]{y15x2, y3-5x2method}. 

In this paper we are particularly interested in using galaxy-CMB cross-correlations for cosmological analyses alone independent of \threetwo{} (i.e. the two probes in \fivetwo{} that are not in \threetwo{}). We are interested in studying these two probes alone for two reasons. First, unlike galaxy-galaxy lensing, the CMB dataset is completely independent from the galaxy tracers in terms of the data acquisition and processing, making it immune to systematic effects that could simultaneously contaminate galaxy position and shear measurements (e.g. depth variation over the footprint). This then provides a robust cross check to the \threetwo{} results. Second, CMB lensing is sensitive to the matter distribution over a wide kernel that peaks around a redshift of 2, while typical galaxy lensing surveys have a sensitivity that peaks at redshift around 1. The cross-correlation thus allows us to measure the large-scale structure at a somewhat higher redshift regime. These two features are particularly interesting given the mild tension seen between ongoing galaxy lensing and CMB analyses, where the galaxy measurements systematically prefer a lower amount of structure compared to the primary CMB constraints \citep{Heymans2021,y3-3x2ptkp}. Recent analysis in \citet{y3-nkgkmeasurement} illustrates how these two probes are already provide interesting constraints on the amount of structure in the Universe.

With the upcoming, much more powerful datasets from Stage-IV galaxy and CMB experiments such as the Vera C. Rubin Observatory's Legacy Survey of Space and Time\footnote{\url{https://www.lsst.org}} (LSST), the ESA's Euclid mission\footnote{\url{https://www.euclid-ec.org}}, the Roman Space Telescope\footnote{\url{https://roman.gsfc.nasa.gov}}, the Simons Observatory\footnote{\url{https://simonsobservatory.org/}} (SO), and CMB Stage-4\footnote{\url{https://cmb-s4.org/}} (CMB-S4), we expect these galaxy-CMB lensing cross-correlation probes will become increasingly important in helping us understand systematic effects associated with individual datasets (e.g. uncertainties in photometric redshift and shear estimation, foreground cleaning). A more in-depth study is therefore warranted for the cross-correlation probes on their own. Here, we are interested in studying systematically the constraining power of the galaxy-CMB lensing cross-correlations as we transition from Stage-III to Stage-IV surveys -- how we bridge our intuition from ongoing datasets to forecasts of the next generation data. Needless to say, our study of the two cross-correlation probes will naturally have implications for 5$\times$2pt and 6$\times$2pt analyses.

We perform simulated likelihood analyses to produce forecasts for Stage-III and Stage-IV experiments in the near and far future. We have chosen DES and {\it Planck} to represent immediately available Stage-III datasets and LSST and SO to represent future Stage-IV data combinations. However, our analysis is fairly general and could be used to guide any Stage-III or Stage-IV analyses. We begin with a ``baseline'' setup of four data combinations (DES$\times${\it Planck}, DES$\times$SO, LSST Y1$\times$SO, LSST Y10$\times$SO). The baseline cases assume galaxy samples that are the same as that used in \threetwo{} analyses, which we draw inspiration from existing DES data and the LSST DESC Science Requirements Document \citep[SRD,][]{SRD2018}. Next, we check whether the baseline setup is robust by trying to reproduce characteristics of the sample using a set of mock galaxy catalogs from LSST Dark Energy Science Collaboration (DESC), \texttt{CosmoDC2} \citep{Korytov2019, Kovacs2022}, with more realistic galaxy properties and measurement uncertainties. Once we have verified the baseline case with \texttt{CosmoDC2}, we then study how deviations from the baseline lens galaxy samples affect the constraints -- we explore variations in the limiting magnitude and maximum redshift range. We only consider magnitude-limited lens samples here as it allows us to systematically study a range of sample variations, and our results could easily be generalized/extrapolated to other more specific samples \citep[see e.g.][]{Tanoglidis2020}. Note that throughout this study, we follow the SRD specification for the source galaxy sample and do not explore the same level of variations as with the lens sample. This is because there is generally less freedom in defining the source sample -- current surveys tend to use every source galaxy for which a good shear measurement can be obtained (e.g. passing certain signal-to-noise and size cuts). In addition to using the LSST DESC simulations, this work also utilizes and facilitates the development of other DESC software including a likelihood code (\textsc{Firecrown})\footnote{\url{https://github.com/LSSTDESC/firecrown}}, a covariance code (\textsc{TJPCov})\footnote{\url{https://github.com/LSSTDESC/TJPCov}}, and a measurement pipeline (\textsc{TXPipe})\footnote{\url{https://github.com/LSSTDESC/TXPipe}}.   

Several previous studies have looked into forecasting the Stage-IV galaxy-CMB lensing cross-correlation. \citet{Schaan:2017} performed a forecast for a $\sixtwo$  analysis for the final LSST and CMB-S4 dataset, highlighting the power of using the cross-correlation probes to self-calibrate nuisance parameters. \citet{Schmittfull2018} focused on using CMB-S4 lensing and LSST galaxy clustering to predict constraints on $\sigma_{8}(z)$ and $f_{\rm NL}$.  SDSS and DESI spectroscopic galaxy samples were also used for comparison. \citet{Chen2021} looked at constraints on $f_{\rm NL}$ and the sum of neutrino mass combining all galaxy-CMB lensing two-point functions between LSST and CMB-S4. \citet{Mishra-Sharma2018} explored the final LSST \threetwo{} combined with either {\it Planck} or CMB-S4 to predict constraints on neutrino mass. \citet{Yu2021} combined CMB-S4 lensing, LSST galaxy clustering and DESI BAO to predict constraints on the dark energy equation of state and neutrino mass. Finally, \citet{Euclid2021} performed similar forecasts for the Euclid mission. The main difference of our work is that we focus on the two cross-correlation probes on their own (galaxy density$\times$CMB lensing and galaxy shear$\times$CMB lensing) and attempt to ground the forecasting work as much as possible to the Stage-III analysis that is currently underway \citep{y3-5x2key}. We incorporate many analysis choices based on Stage-III analyses, and study the increment change between the end of Stage-III to the beginning of Stage-IV to understand what factors contribute to qualitative changes in the Stage-IV forecasts compared to Stage-III. Recent work from \citet{Fang2021} explores similar questions, but focuses more on the power of using the same source and lens sample in a full $\sixtwo{}$ analysis.

The outline of the paper is as follows. In Section~\ref{sec:theory} we describe the formalism used to model the galaxy-CMB lensing cross-correlation, for both the cosmological signal and the systematic effects, which is used for both the simulated data vector and analytic covariance. In Section~\ref{sec:modeling_samples} we describe the procedure we take to model the different datasets both for galaxy and CMB surveys, including models of the nuisance parameters extracted from mock galaxy catalogs. These data characteristics will be relevant for constructing the analytic covariance matrix. In Section~\ref{sec:analysis} we describe several important components in our analysis including the covariance matrix, the cosmological inference software and the figure of merit we use to quantify the constraining power of each data combination. In Section~\ref{sec:results} we present our results of the projection for the cosmological constraints of the galaxy-CMB lensing cross-correlation probes going from Stage-III to Stage-IV datasets, first with the baseline galaxy samples in the LSST DESC SRD, then with variations over the baseline. We summarize our results in Section~\ref{sec:summary}. 

\section{Modeling}
\label{sec:theory}
In this section, we discuss relevant theoretical models for the cross correlations, noise in the CMB lensing map, and astrophysical/observational systematic effects. This section also lays the foundation for generating analytic covariance in Section~\ref{sec:cov}. 

\subsection{Galaxy-CMB lensing cross-correlation}

In this work we focus on two cross-correlation probes between three cosmic fields: galaxy density ($\delta_{g}$), galaxy weak lensing shear ($\gamma$) and CMB lensing convergence ($\kappa_{\rm CMB}$). The two cross-correlations are: galaxy density $\times$ CMB lensing convergence \nk{} and galaxy weak lensing shear $\times$ CMB lensing convergence \gk{}. In this work we will model the observable in configuration space similar to \citet{baxter2019}. Following standard convention, we will refer to the galaxy density tracers as the {\it lens galaxies} and the galaxies used to measure the weak lensing signal as the {\it source galaxies}.

To derive predictions for the cross-correlation function, we start by writing down the cross power spectra in harmonic space for a multipole $\ell$. Using the Limber approximation, we have
\begin{equation}
C_{\delta_{g}^{i} \kappa_{CMB}}(\ell)  = \\
\int d\chi \frac{q^i_{\delta_g} \left(\chi \right) q_{\kappa_{\rm CMB}} (\chi) }{\chi^2} P_{\rm NL} \left( \frac{\ell+1/2}{\chi}, z(\chi)\right),
\label{eq:CNK}
\end{equation}
and 
\begin{equation}
C_{\gamma^{i} \kappa_{CMB}}(\ell) = \\\int d\chi \frac{q^i_{\gamma} (\chi) q_{\kappa_{\rm CMB}} (\chi) }{\chi^2} P_{\rm NL} \left( \frac{\ell+1/2}{\chi}, z(\chi)\right),
\label{eq:CGK}
\end{equation}
where $i$ labels the redshift bin (of either the lens or source galaxies), $\chi$ is the comoving distance along the line-of-sight and $P_{\rm NL}(k,z)$ is the nonlinear matter power spectrum. We compute the nonlinear power spectrum using the Boltzmann code CAMB\footnote{See \texttt{camb.info}.} \citep{Lewis:2000,Howlett:2012} with the Halofit extension to nonlinear scales \citep{Smith:2003,Takahashi:2012} and the \citet{Bird:2002} neutrino extension. For each of the tracers ($\delta_{g}$, $\gamma$ and $\kappa_{\rm CMB}$), there is an associated weight (or ``kernel'') $q(\chi)$ that encodes the sensitivity of the tracer to structure along the line-of-sight. For sources at comoving distance $\chi$, this weight is most sensitive approximately halfway between the observer and the source, and is given by
\begin{equation}
q_{\gamma}^i (\chi) = \frac{3 \Omega_m H_0^2}{2c^2}\frac{\chi}{a(\chi)}\int_{\chi}^{\infty} d\chi' n^i_s(z(\chi')) \frac{dz}{d\chi'} \frac{\chi' - \chi}{\chi'},
\label{eq:weight_kappa}
\end{equation}
where $n_s(z)$ the normalized number density of the source galaxies as a function of redshift, $H_0$ and $\Omega_{m}$ are the Hubble constant and matter density parameters, respectively. $a(\chi)$ is the scale factor corresponding to comoving distance $\chi$. The CMB lensing weight is
\begin{equation}
q_{\kappa_{\rm CMB}} (\chi) = \frac{3 \Omega_{\rm m} H_0^2}{2c^2}\frac{\chi}{a(\chi)}  \frac{\chi^* - \chi}{\chi^*}, \label{eq:weight_cmbkappa}
\end{equation}
where $\chi^*$ denotes the comoving distance to the surface of last scatter. For galaxy density, the weight is 
\begin{equation}
q_{\delta_g}^i (\chi) = b_g^i(\chi) n_g^i(z(\chi)) \frac{dz}{d\chi}, \label{eq:weight_gal}
\end{equation}
where $n_g^i(z)$ is the normalized number density of the lens galaxies in the $i$th bin as a function of redshift. We will further simplify the bias modeling such that the bias for each galaxy redshift bin is assumed to be a constant, $b_g^i$.  In reality, the linear bias model is known to break down at small scales \citep{Zehavi2005,Blanton2006,Cresswell2009}. Following previous work from \citet{baxter2019}, we choose a set of scale cuts that mitigate the bias caused by the poor assumption of linear bias on small scales (see Section~\ref{sec:cov}). We note that we have ignored lensing magnification in our model, though if we follow the approach taken in e.g. \citet{y3-5x2method}, we do not expect it to change the cosmological constraints significantly since there is no additional free parameters involved.

Since CMB experiments have finite-size beams, when generating the $\kappa_{\rm CMB}$ map this beam is deconvolved, exponentially increasing noise at small scales. A small amount of smoothing is applied to the map to suppress numerical instabilities in the covariance due to this noise. We use a Gaussian smoothing with full width at half maximum of $\theta_{\rm FWHM}$. In harmonic space, this corresponds to multiplication of the maps by
\begin{equation}
B(\ell) = \exp (-\ell(\ell + 1)/\ell_{\rm smooth}^2),
\end{equation}
where $\ell_{\rm smooth} \equiv \sqrt{16 \ln 2}/\theta_{\rm FWHM}$.  Additionally, we filter out modes in the $\kappa_{\rm CMB}$ map with $\ell < \ell_{\rm min}$ and $\ell > \ell_{\rm max}$, where the lower bound is to avoid the potential contamination coming from the mean-field calibration in the CMB lensing map and the upper limit is imposed to remove potential biases due to foregrounds in the $\kappa_{\rm CMB}$ map.

Converting the above expressions to configuration-space correlation functions under the flat-sky approximation via a Legendre transform yields
\begin{eqnarray}
\langle \gamma \kappa_{\rm CMB} \rangle &=& \int \frac{d\ell \, \ell}{2\pi} F(\ell) J_2(\ell \theta) C_{\gamma \kappa_{\rm CMB}}(\ell) ,\label{eq:shearkcorr} \\
\langle \delta_{g} \kappa_{\rm CMB} \rangle &=& \int \frac{d\ell \, \ell}{2\pi} F(\ell) J_0(\ell \theta) C_{\delta_{g} \kappa_{\rm CMB}}(\ell),
\label{eq:galkcorr}
\end{eqnarray}
where $J_0$/$J_2$ is the zeroth/second order Bessel function of the first kind. 
The appearance of $J_2$ in Eq.~\ref{eq:shearkcorr} is a consequence of our decision to measure the correlation of $\kappa_{\rm CMB}$ with tangential shear \citep[similar to][]{baxter2019}. The function $F(\ell) = B(\ell) \Theta(\ell - \ell_{\rm min}) \Theta(\ell_{\rm max} - \ell)$, where $\Theta(\ell)$ is a step function, describes the filtering that is applied to the $\kappa_{\rm CMB}$ map.

\subsection{Nuisance parameters}
\label{sec:nuisance_parameters}
Our model includes the following nuisance parameters:
\begin{itemize}
    \item {\bf Galaxy bias:} As discussed above, we use a linear galaxy bias that is assumed to be constant in a given lens redshift bin $i$ and across angular scales. Or 
    \begin{equation}
        b^{i}_{g}(\chi, \theta) = b^{i}.
    \end{equation}
    \item {\bf Uncertainty in redshift distribution estimation:} Following a parametrization commonly used in galaxy surveys \citep[e.g.][]{Krause:2021}, we assume the uncertainty in the redshift distribution estimation can be modeled via a shift in the mean. Or 
    \begin{equation}
    \label{eq:nz}
    n_{i}(z) \rightarrow n_{i}(z+\Delta^{i}_{z}),
    \end{equation}
    where $i$ represents the redshift bin. Note that the same model is used for both the source and the lens samples but with different priors depending on the sample.
    \item {\bf Uncertainty in shear estimation:} Following a parametrization commonly used in galaxy surveys \citep[e.g.][]{Krause:2017}, we assume the uncertainty in the shear estimation can be modeled via a multiplicative factor that is constant in each source redshift bin $i$ and scale-independent. Or
    \begin{equation}\label{eq:shear_bias}
    C_{\gamma^{i}\kappa_{\rm CMB}}(\ell) \rightarrow (1+m^{i})C_{\gamma^{i}\kappa_{\rm CMB}}(\ell),\nonumber
    \end{equation}
        
    \item {\bf Intrinsic alignment:} Intrinsic alignment (IA) refers to the fact that even in the absence of lensing, galaxies could have preferred orientations depending on their environment and formation history \citep[see e.g.][]{blazek2019}. We adopt a two-parameter Nonlinear Linear Alignment model \citep[NLA,][]{bridle07} for IA, which effectively carries out the following replacement.
\begin{equation}
q_{\rm \gamma^i}(\chi) \rightarrow q_{\gamma^i}(\chi) - A(z(\chi)) \frac{n_{\rm s}^i(z(\chi))}{\bar{n}_{\rm s}^i} \frac{dz}{d\chi},
\end{equation}
where
\begin{equation}
A(z) = A_{0} \left( \frac{1+z}{1+z_0} \right)^{\eta_{\rm IA}} \frac{0.0139\Omega_{\rm m} }{G(z)},
\end{equation}
and where $G(z)$ is the linear growth factor and $z_0$ is the redshift pivot point which is usually set to the mean redshift of the source sample. Here we set it to $0.62$ throughout, which is close to the mean redshift for DES sources, though we check that our results do not change when setting it to the actual mean redshift for each sample. The two free parameters in this model are $A_{0}$ and $\eta_{\rm IA}$.
\end{itemize}

\section{Modeling the Dataset}
\label{sec:modeling_samples}

The goal of this study is to systematically explore how the cosmological constraints change for galaxy-CMB lensing cross-correlation when different galaxy samples are used. We will focus on dataset combinations that will become available in the near and far future. In particular, on the galaxy survey side we explore three cases: the final six-year dataset expected from the Dark Energy Survey (DES), the first year data from the Rubin Observatory's Legacy Survey of Space and Time (LSST), and the final ten-year LSST data. On the CMB side, we explore two cases: the legacy {\it Planck} dataset and the final dataset expected from the Simon Observatory (SO). For concreteness, we will discuss four data combinations: 
\begin{enumerate}
    \item[(i)] {\bf DES $\times$ {\textit{\textbf{Planck}}}:} This scenario represents an analysis that could be carried out in the near future with upcoming DES releases and existing CMB data. 
    \item[(ii)] {\bf DES $\times$ SO:} This scenario represents what could be achieved as the new CMB datasets become available. Comparing with the previous case also gives us information of what improving only the CMB datasets will add.
    \item[(iii)] {\bf LSST Y1$\times$SO:} This scenario represents what could be expected when LSST Y1 data is available. Comparing with the previous case also gives us information of what improving only the galaxy datasets will add.
    \item[(iv)] {\bf LSST Y10$\times$SO:} This final scenario represents the ultimate galaxy-CMB lensing dataset we will be able to analyze in the coming decade.
\end{enumerate}
We note that these cases were chosen to best illustrate how the different galaxy samples interact with the cosmological constraints. Not all of them are particularly realistic in terms of time scale. For example, SO data may not be available for the second and third scenarios, but fixing the CMB dataset and varying the galaxy sample in the last three cases allow us to more clearly understand the improvement coming from the galaxy sample.

In practice, the different CMB datasets are modeled through different noise power spectra and filtering functions. On the other hand, the different galaxy datasets, as well as the different galaxy samples within the same dataset, correspond to different redshift distributions, redshift errors, galaxy bias, shape noise and galaxy number density. These data characteristics from both the CMB and galaxy datasets are fed into both the data vectors and the covariance matrix. In what follows, we briefly describe how we model the different datasets as a whole for the CMB lensing maps (Section~\ref{sec:cmb_maps}) and for the galaxy samples (Section~\ref{sec:galaxy}). Next we discuss in more detail how we use mock galaxy catalogs to extract models for nuisance parameters and $n(z)$ for different galaxy samples (Section~\ref{sec:samples}).

\subsection{CMB lensing maps} 
\label{sec:cmb_maps}

\begin{figure}
    \centering
    \includegraphics[width=0.9\linewidth]{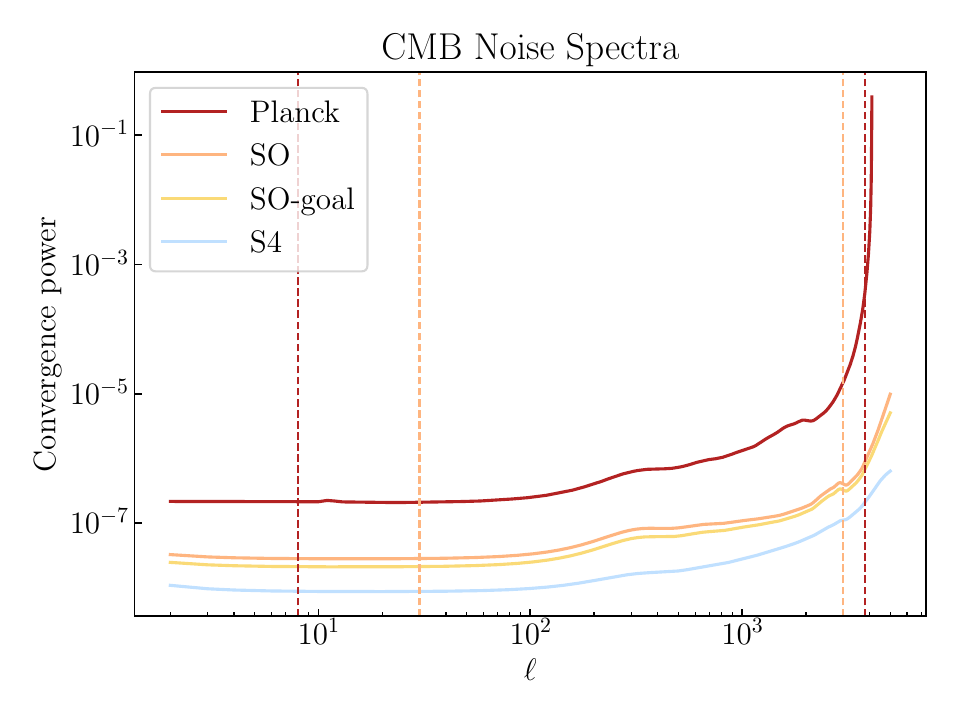}
    \caption{Noise power spectrum for the CMB lensing maps used in this work. The dashed vertical lines indicate the $\ell_{\rm min}$ and $\ell_{\rm max}$ cut applied to the CMB convergence maps. This work uses \textit{Planck} and SO. SO-goal and S4 curves are used to illustrate more optimistic estimates.}
    \label{fig:cmb_noise}
\end{figure}

We consider the {\it Planck} CMB lensing map described in \cite{Planck2020} to represent the ``current'' wide-field CMB lensing map that overlaps fully with DES and is therefore ideal for cross-correlation studies. This map uses both temperature and polarization data from {\it Planck} and covers about 70\% of the sky. The noise power spectrum associated with the lensing map \footnote{\texttt{MV/nlkk.dat} in \texttt{COM\_Lensing\_4096\_R3.00.tgz}, taken from \url{https://pla.esac.esa.int/\#home}} is shown in Figure~\ref{fig:cmb_noise}. To model the CMB lensing signal and associated covariance, we assume the map to be filtered similar to that done in \citet{y3-5x2method}, which involves $\ell_{\rm min}=8$, $\ell_{\rm max}=3800$ and a Gaussian smoothing of FWHM$=$8 arcmin. 

The Simons Observatory (SO) is a new CMB experiment being built on Cerro Toco in Chile, due to begin observations by 2023. SO will measure the temperature and polarization anisotropy of the CMB in six frequency bands (27, 39, 93, 145, 225 and 280 GHz). The survey will map $\sim$40\% of the sky, overlapping with the majority of the LSST footprint. For our forecast, we use their {\it baseline} projected lensing noise power spectrum\footnote{Taken from \url{https://github.com/sriniraghunathan/ilc/blob/master/results/lensing_noise_curves/sobaseline_lmin100_lmax5000_lmaxtt3000.npy}}. For modeling, we assume $\ell_{\rm min}=30$ and $\ell_{\rm max}=3000$ filtering following \citet{Ade2019}\footnote{We have checked that our results are not very sensitive to the exact choice of $\ell_{\rm min}$ and $\ell_{\rm max}$.}. In addition, the maps are smoothed by a Gaussian filter of FWHM$=$1 arcmin (approximate size of the SO beam). The noise power spectra are shown in Figure~\ref{fig:cmb_noise}, labeled ``SO''. For comparison, we also show two other noise curves: a more optimistic scenario for SO labelled ``SO-goal'' and one for CMB-S4. The results from this paper could easily be extrapolated to these more lower noise datasets.  

\subsection{Galaxy samples}
\label{sec:galaxy}

For galaxy samples, we specify the source and lens samples separately. For both samples, we model the underlying redshift distribution via a parametrized model 
\begin{equation}
    \frac{dN}{dz}\propto z^2\exp[-(z/z_0)^\alpha],
\label{eq:nz}
\end{equation}
where the parameters $(z_0, \alpha)$, whose values can be found in Table~\ref{tab:SRD}, are taken from the LSST DESC SRD and were derived from fitting to the LSST mock galaxy catalog \textsc{CatSim}\footnote{\url{https://www.lsst.org/scientists/simulations/catsim}} and checked with the Deep2 spectroscopic survey \citep{Newman2013}. To form the tomographic bins, we apply top-hat window functions to the analytical redshift distribution, then convolve with a Gaussian filter with width $\sigma_z$ (referred to as the \textit{redshift error} hereafter). We note that this is an approximate treatment that mimics the effect of tomographic bins with photometric redshift. We explore more realistic redshift distributions in the next section. 

The lens sample is characterized by a given magnitude selection and associated number density ($n_l$) and galaxy bias ($b_{\rm gal}$). We then use one parameter to capture the uncertainty in the mean redshift ($\sigma_{z_{b}}$). The source sample is characterized by its number density ($n_{s}$) and shape noise ($\sigma_e$). Shape noise refers to the intrinsic galaxy shapes that do not contain cosmological signal. In weak lensing, the observed galaxy shape is a combination of cosmological shear $\gamma$ and shape noise. Here the shape noise is characterized by the standard deviation of the single-component ellipticity distribution  $\sigma_e$. We then use one parameter to capture the uncertainty in the mean redshift ($\sigma_{z_{b}}$) and the uncertainty in the shear calibration ($\sigma_{\rm m}$).

We describe below our approach to model these samples for the baseline galaxy datasets.

\subsubsection{DES}
\label{sec:des}

DES \citep{DES2005} covers 5,000 deg$^{2}$ with five optical filter bands ($grizY$). The imaging data is taken with the 570-million pixel Dark Energy Camera \citep[DECam;][]{Flaugher2015} at the 4$\rm m$ Blanco telescope at the Cerro Tololo Inter-American Observatory, with a field of view of $\sim2$deg$^{2}$. Data collection for the survey started in Fall of 2014 and ended early 2019. The final dataset reaches a depth of $i\sim23.7$ across the footprint.    

Our model for the baseline DES galaxy sample and modeling choices are largely informed by the existing DES Y3 analysis of cross-correlation with the CMB lensing map from {\it Planck} and the South Pole Telescope (SPT) \citep{y3-5x2key}, but we extrapolate to the final (six-year, or Y6) dataset whenever appropriate. The full DES (Y6) dataset will nearly double the total exposure time of the DES Y3 data and is expected to be slightly shallower than LSST Y1, and with a factor of 3-4 smaller sky area. We assume 5 lens galaxy bins with number density similar to the sample developed in \cite{y3-2x2maglimforecast}. This sample uses a redshift-dependent magnitude cut ($i<4 z_{\rm mean} + 18$, where $z_{\rm mean}$ is the mean redshift in that bin), which was designed to select bright galaxies of roughly the same number density across the redshift bins. This yields a total lens galaxy number density about 1/arcmin$^2$. The values for galaxy bias, redshift error and bias in mean redshift used in this work are chosen to be roughly similar to that used in the DES Y3 data. Here we have assumed that the final DES analysis adopts the same lens sample as DES Y3, though see Section~\ref{sec:maglim} for an alternative choice. For the source galaxy sample, we again  mimic the sample in DES, assuming 4 tomographic bins between $z=0.2$ and 1.3, with a shape noise of 0.26 and number density $\sim 50\%$ more compared to the DES Y3 shear catalog \citep{y3-shapecatalog}. The uncertainty in the shear calibration is also informed by DES Y3 studies \citep{y3-imagesims}. Due to the similar survey depth of DES and LSST Y1, we assume the same underlying lens and source redshift distributions for both, which we use the parametric form described in the next section describing LSST.

\begin{figure*}
    \centering
    \includegraphics[width=\linewidth]{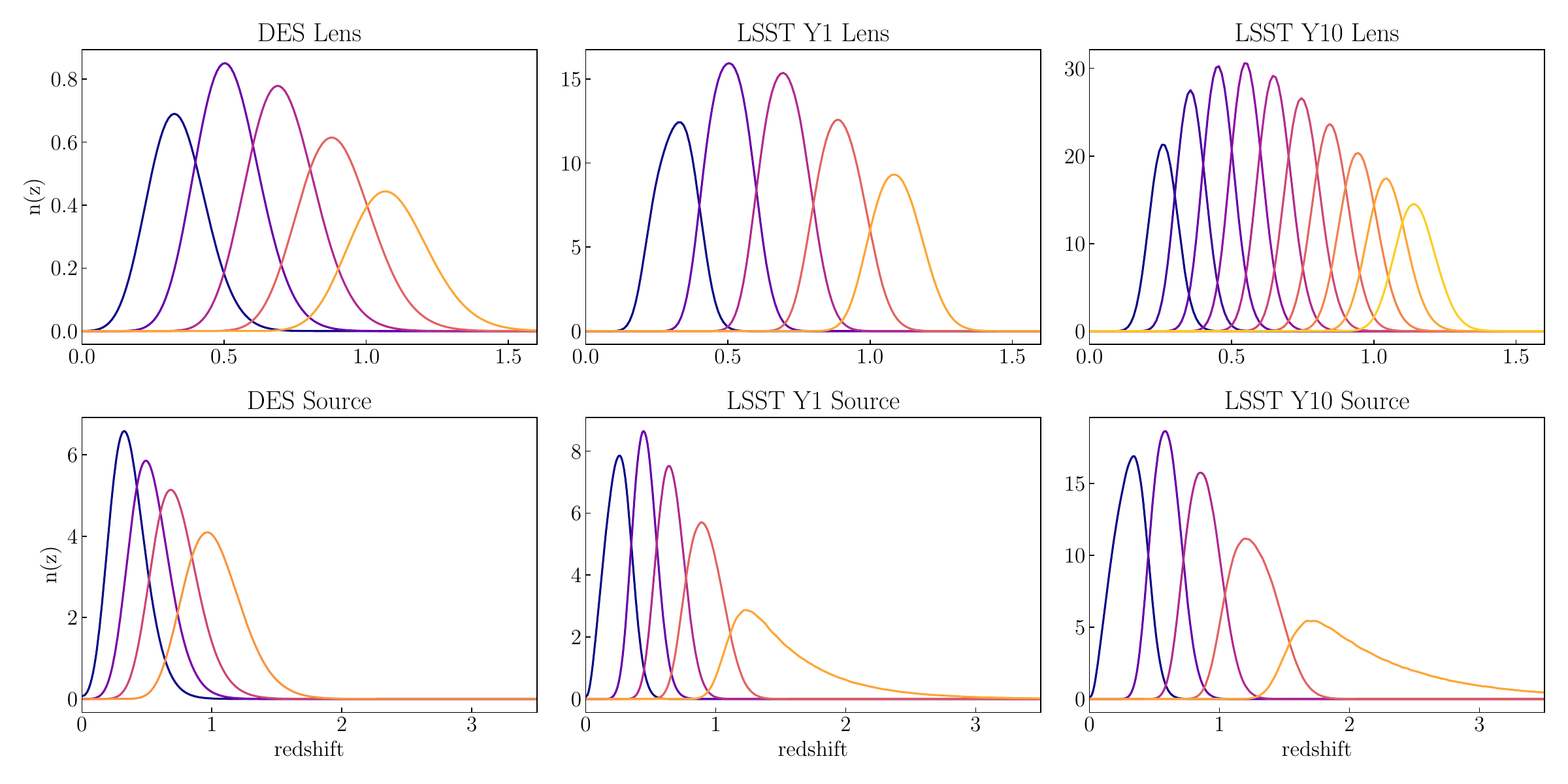}
    \caption{Lens (upper panels) and source (lower panels) sample redshift distributions for each tomographic bin in the baseline setup for DES (left), LSST Y1 (middle) and LSST Y10 (right). The y-axis shows the number density (arcmin$^{-2}$) in a $\Delta z=1$ interval.}
    \label{fig:Y1_fiducial_nz}
\end{figure*}

We list in Table~\ref{tab:SRD} the basic characteristics of the DES galaxy samples we assume for our analysis. The redshift distributions for the fiducial lens sample are shown in Figure~\ref{fig:Y1_fiducial_nz}. We note that unlike the LSST data characteristics described in the next section, which are projections, the DES numbers are informed by a mix of data that was already taken, as well as conservative extrapolations. 

\subsubsection{LSST}
\label{sec:lsst}

LSST \citep{Abell2009} is a Stage-IV optical ground-based galaxy survey that is scheduled to begin commissioning late 2021 and start its 10-year science survey in 2023. The survey will use six filter bands ($ugrizY$) to survey $\sim 18,000$ deg$^{2}$ in the Southern hemisphere to a depth of $i \sim 26.8$.

Our fiducial models of the LSST Y1 and Y10 data rely heavily on the LSST DESC SRD. The SRD specifies the data to be delivered and requirements to be met by DESC analyses, which include a description of the galaxy samples used for DESC's flagship \threetwo{} analysis. In particular, we use the ``Gold'' sample in the SRD to serve as our fiducial sample for LSST and explore variations from the Gold sample. In the SRD, the footprints for LSST Y1 and Y10 are 12,300 deg$^2$ and 14,300 deg$^2$ respectively. These areas exclude the region of the LSST that is close to the Galactic plane and has high extinction. For the lens sample, LSST Y1 specifies 5 tomographic bins between $z=0.2$ and 1.2, while LSST Y10 increases the number of bins to 10 with the same overall range. The lens selection adopts a simple magnitude cut of $i<24.1$ (LSST Y1) and $i<25.3$ (LSST Y10), reflecting the increased depth of the dataset. This yields a total galaxy number density of 13.6/arcmin$^{2}$ and 35.7/arcmin$^{2}$ -- we note that this is significantly higher than the DES selection. The galaxy bias and redshift error values are taken from SRD. For the source sample, we use 5 tomographic bins with equal number of galaxies in each bin. The shape noise and redshift error values are taken from SRD. The uncertainty in the mean redshift for both the lens and source galaxies, as well as the uncertainty in the shear calibration are set to the {\it requirements} in SRD as opposed to an estimate. That is, the value needed so that the dark energy constraints are not systematics-limited. The main rationale is that we expect significant methodology development in the area of redshift and shear calibration leading up to LSST and it is challenging to forecast these values from first principle.

\begin{table*}
    \centering
    \begin{threeparttable}
    \begin{tabular}[htpb]{c|c|c|c}
    & DES & LSST Y1    & LSST Y10  \\
    & (5,000 deg$^2$) & (12,300 deg$^2$) & (14,300 deg$^2$) \\
    \hline
    \multicolumn{4}{l}{Lens sample} \\
    \hline
    Redshift binning    & $0.2<z<1.2$ &  $0.2<z<1.2$ &$0.2<z<1.2$\\
    &5 bins ($\Delta z=0.2$) & 5 bins ($\Delta z=0.2$) & 10 bins ($\Delta z=0.1$)\\
    Limiting magnitude $m_{\rm lim}$ & $4 z_{\rm mean} + 18$ & $24.1$ & $25.3$\\
    Redshift distribution ($z_0,\alpha$) & (0.26, 0.94) & (0.26, 0.94) & (0.24,0.90)\\
    Number density normalization $\hat{n}_{l}$ & 1.3/arcmin$^2$ & 18/arcmin$^2$  & 48/arcmin$^2$\\
    Number density $n_{l}$ & 1/arcmin$^2$ & 13.6/arcmin$^2$  & 35.7/arcmin$^2$\\
    Redshift error $\sigma_{z}$ & $0.06(1 + z)$ & $0.03(1 + z)$ & $0.03(1 + z)$\\
    Uncertainty in mean redshift $\sigma_{z_b}$ &$0.03(1 + z)$\tnote{a} & $0.005(1 + z)^{*}$ & $0.003(1 + z)^{*}$\\
    Galaxy bias $b_{\rm gal}$& $1.3/G(z)$ & $1.05/G(z)$ & $0.95/G(z)$\\
    Minimum scale cut & 10 arcmin & 10 arcmin  & 10 arcmin \\
    \hline 
    \multicolumn{4}{l}{Source sample} \\
    \hline
    Redshift binning &4 bins; $0.2<z<1.3$& 5 bins & 5 bins \\
      &equal galaxy number & equal galaxy number &  equal galaxy number \\
    Redshift distribution ($z_0,\alpha$) & (0.13, 0.78) & (0.13, 0.78)  & (0.11, 0.68)\\
    Number density $n_s$& 9/arcmin$^2$ & 10/arcmin$^2$  & 27/arcmin$^2$ \\
    Shape noise $\sigma_{e}$& 0.26 & 0.26 & 0.26\\
    Redshift error $\sigma_{z}$ &$0.1(1 + z)$ & $0.05(1 + z)$ & $0.05(1 + z)$\\
    Uncertainty in mean redshift $\sigma_{z_b}$ &$0.01(1 + z)$& $0.002(1 + z)^{*}$ & $0.001(1 + z)^{*}$\\
    Uncertainty in shear calibration $\sigma_{\rm m}$ & $0.01$ & $0.013^{*}$ & $0.003^{*}$\\
    Minimum scale cut & 5 arcmin & 5 arcmin & 5 arcmin \\
    \end{tabular}

    \caption[Caption for LOF]{DES, LSST Y1 and Y10 source and lens sample specifications in our baseline setup. These are based on a combination of existing DES Y3 data plus extrapolations for DES Y6, and the LSST DESC SRD. The items with a superscript $*$ indicates these are SRD {\it requirements}\tnote{b}. The magnitude selection is applied on the $i$ band magnitude, or $i<m_{\rm lim}$. The redshift distributions follow a parametrized model in Equation~\ref{eq:nz}. For the lenses, we list both the number density normalization $\hat{n}_{l}$, which is the total number density if one were to integrate over the full redshift distribution ignoring the tomographic bins, and the actual number density $n_{l}$ -- the former is what is listed in the LSST SRD.}

    \begin{tablenotes}
    \item [a] 
    We note that after the publication of this work we discovered that this value is overestimated. A more reasonable value based on DES Y3 is $\sim 0.006$ \citep{y3-2x2maglimforecast}. However, as the constraining power from \nkgk{} is already limited by cosmic variance, this overestimation should not qualitatively change our results.
    \item [b] While we try to keep our analysis choices as close to those of the LSST DESC official forecast as possible, the official forecast numbers for these items are not available in the LSST SRD, only \textit{requirements} of these numbers are specified.
    \end{tablenotes}
    \label{tab:SRD}
    \end{threeparttable}
\end{table*}

We list in Table~\ref{tab:SRD} the basic characteristics of the galaxy samples we assume for our analysis. The redshift distributions for the fiducial lens samples are shown in Figure~\ref{fig:Y1_fiducial_nz}. A few observations can be made: First, as we already hinted, the lens galaxy sample is significantly larger in LSST compared to DES. This is partially due to the fairly optimistic selection used in the LSST DESC SRD -- it is not entirely clear from e.g. \cite{y3-2x2maglimforecast} that one would be able to properly characterize the redshift distribution for a sample as faint as $i<25.3$. Second, the uncertainties on the mean of the photometric redshift $\sigma_{z_{b}}$ changes by a factor of 5-6 (10) from DES to LSST Y1 (Y10). These improvements could be considered fairly optimistic especially for LSST Y1, and as we will discuss later, contributes to some of the main findings of this work. Finally, for the source galaxy sample, the assumption that there is no upper bound in the redshift bins yields a very high redshift tail in the LSST sources that is not entirely realistic -- characterizing the last redshift bin could be extremely challenging given the lack of any calibration data (either spectroscopic samples or luminous red galaxies with well-known redshifts), and it is especially unlikely that we will reach the level of uncertainty requirements in Table~\ref{tab:SRD} for these high-redshift sources. 

\subsection{Modeling variant samples with \texttt{CosmoDC2}} 
\label{sec:samples}

In addition to the fiducial samples, we explore variations on the lens samples in their limiting magnitude and redshift range -- these two are the primary characteristics that define the lens samples. We select the variant samples in the following way:
\begin{itemize}
    \item {\bf Limiting magnitude:} for each dataset, we start with their respective baseline limiting magnitude selection in Table~\ref{tab:SRD}, then we increase and decrease the limiting magnitude by 0.5 and 1.  
    \item {\bf Redshift range:} for each dataset, we start with the baseline binning scheme and add or remove bins so the maximum redshift range increases or decreases by 0.2 and 0.4 (that is, for DES and LSST Y1, we add/remove 1 or 2 bins, and for LSST Y10, we add/remove 2 or 4 bins).
\end{itemize}
To estimate the sample characteristics of the variant lens samples, we appeal to mock galaxy catalogs. In particular, we use the LSST DESC \texttt{CosmoDC2} catalogs \citep{Korytov2019,Kovacs2022}. \texttt{CosmoDC2} is a mock galaxy catalog based on the Outers Rim N-body simulation \citep{Heitmann2019}. It covers $\sim400$ deg$^{2}$ area to the redshift and depth much beyond that expected for LSST. Each galaxy is assigned observable quantities from photometry to morphology based on a number of semi-analytical models. The galaxies also carry cosmological information from both their position on the sky and ray-traced weak lensing properties. These observable quantities allow us to generate well-defined samples similar to what is done in observational data, while knowing exactly what the expected cosmological signal is. To derive the galaxy bias in Section~\ref{sec:galaxy_bias} we generate an octant-sized CMB lensing convergence map from the Outer Rim particle lightcones. This makes use of the density shells detailed in \citet{Korytov2019}, alongside a Gaussian random field at $z>3$, applying Wigner filtering using the \textsc{Lenspix} code of \citet{lenspix}, and summing the contribution of the lens planes under the Born approximation.

With \texttt{CosmoDC2} we are then able to carry out different selections on the lens sample and build models for how the redshift distribution, galaxy number density, galaxy bias, redshift error, and uncertainty on the mean redshift change. Below we describe our approach for extracting the different model components from \texttt{CosmoDC2}. We note that this work primarily focuses on the lens sample selection, thus we will fix the source sample to be the same as the baseline case listed in Table~\ref{tab:SRD}.

\subsubsection{Redshift distribution}
\label{sec:nz_cosmodc2}

The redshift distribution of the lens galaxies ($n_{g}(z)$ in Equation~\ref{eq:weight_gal}) is the true redshift distribution of galaxies in a given tomographic bin. Operationally, the bins are often selected via noisy quantities such as single-point photo-$z$ estimates, and the $n(z)$ estimation is a complicated multi-step procedure in ongoing galaxy survey analyses \citep{y3-sompz,y3-lenswz}. Here we assume the estimation is perfect modulo an error in the mean of the $n(z)$, which we will discuss in Section~\ref{sec:bias_mean_z}. The shape of the $n(z)$'s extracted from \texttt{CosmoDC2} is more realistic than the baseline model since it incorporates photometric errors. After applying any magnitude cuts, we divide the lens galaxies into tomographic bins based on their single-point photo-$z$ estimate (\texttt{photoz\_mean} in \texttt{CosmoDC2}), and then use the distribution for the true redshift as $n(z)$. In the left panel of Figure~\ref{fig:Y1_fiducial_nz_cosmodc2}, we show an example of the lens $n(z)$ distributions from \texttt{CosmoDC2} in the LSST Y1 case compared to the analytic form in the LSST DESC SRD. We note that when extracting the $n(z)$ from the catalogs, we do not need the redshift error ($\sigma_z$) parameter in Table~\ref{tab:SRD} since the $n(z)$ derived above already includes the redshift error captured by $\sigma_z$. In the right panel of Figure~\ref{fig:Y1_fiducial_nz_cosmodc2} we show an example (using the 3rd bin in the left panel) of how $n(z)$ changes when we vary the magnitude limit of the sample. As expected, with a fainter magnitude limit, we have more galaxies in the sample, and the shape of the $n(z)$ also changes, such that the tails of the distribution extend to higher and lower redshifts.

\subsubsection{Galaxy number density}

The number densities of galaxies in each tomographic bin is primarily used in the computation of the covariance matrix (see Section~\ref{sec:cov}). We can estimate it directly from \texttt{CosmoDC2} by counting the number of galaxies after applying the redshift and magnitude cuts described in Table~\ref{tab:SRD}. We note that similar to Section~\ref{sec:nz_cosmodc2}, the redshift selection is done on the single-point photo-$z$ estimate \texttt{photoz\_mean} instead of true redshift.

\subsubsection{Galaxy bias}
\label{sec:galaxy_bias}
The different lens galaxy samples will have different galaxy biases, which is needed in our modeling of the data vector and the covariance. We extract these galaxy bias values from direct measurements on the \texttt{CosmoDC2} catalog. For a given tomographic bin of a given dataset, we measure the cross-correlation of the galaxy density and the noiseless CMB lensing map in \texttt{CosmoDC2}, as well as its jackknife covariance matrix using the software package \texttt{treecorr}\footnote{\url{https://github.com/rmjarvis/TreeCorr}}. Holding all other cosmological parameters at the input values of the \texttt{CosmoDC2} simulation, we fit for a linear galaxy bias at scales $>5$Mpc/$h$. 

One caveat in the bias measurement from \texttt{CosmoDC2} is that, due to the simulation's resolution limit, the galaxies with halo mass $<10^{11}M_\odot$ were added to the catalogs without an association with dark matter halos. These galaxies therefore are distributed randomly and will have a galaxy bias of zero. They can be identified with \texttt{halo\_id}$<0$ and should be removed from the sample to obtain a correct bias. We show in Appendix~\ref{sec:ultrafaint}, Figure~\ref{fig:haloid0} the fraction of galaxies ranging from 31\% to 15\% that are not correlated with dark matter in the LSST Y1 baseline case
-- we remove these galaxies when measuring the galaxy bias. This effectively means we overestimate the galaxy bias slightly especially in the low redshift bins. As we discuss in Appendix~\ref{sec:ultrafaint}, the effect of this error on the resulting cosmological constraints is small.

\begin{figure}
    \centering
    \includegraphics[width=0.49\textwidth]{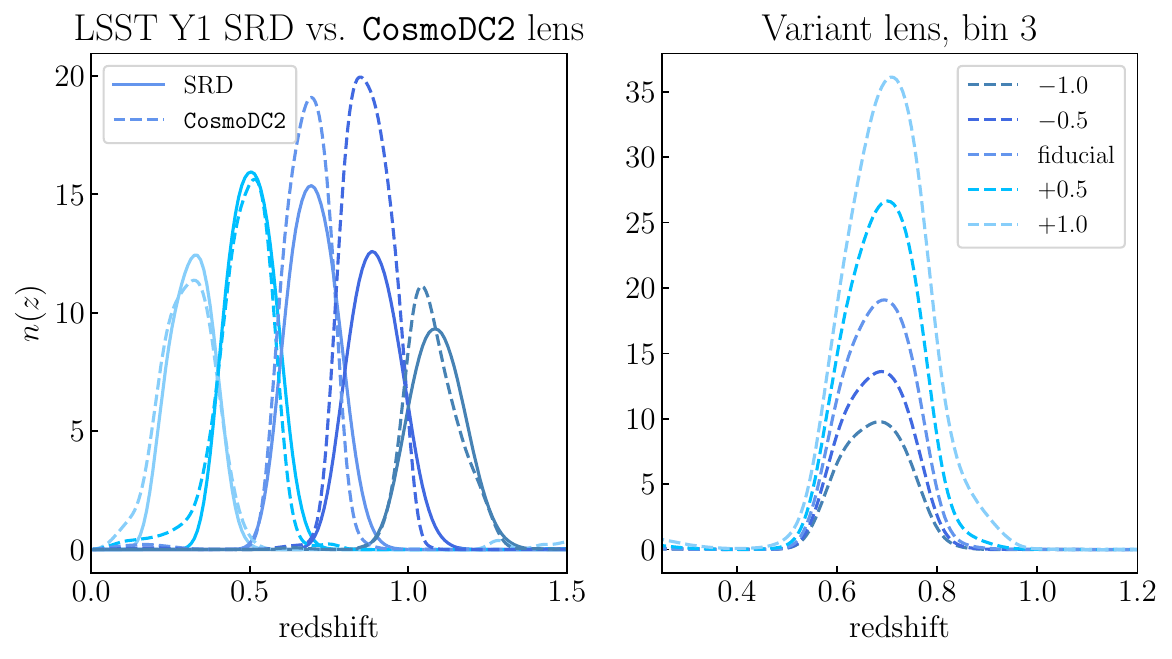}
    \caption{\textit{Left panel}: For LSST Y1, comparison between analytic $n(z)$ given by the SRD (solid) and that extracted from \texttt{CosmoDC2} $n(z)$ (dashed) for the baseline lens samples. \textit{Right panel}: For the third bin in the left panel, how the $n(z)$ changes when we vary the limiting magnitude. The y-axis shows the number density (arcmin$^{-2}$) in a $\Delta z=1$ interval.}
    \label{fig:Y1_fiducial_nz_cosmodc2}
\end{figure}

The direct measurement of galaxy bias from \texttt{CosmoDC2} is rather noisy given the small area coverage. Thus, we follow the LSST DESC SRD convention and assume that the linear galaxy bias takes the form:  
\begin{equation}
    b_{\rm gal}(z) = \frac{b_{\rm gal, 0}}{G(z)},
\end{equation}
where $b_{\rm gal, 0}$ is a coefficient depending on the magnitude limit. For each variant sample, we measure the galaxy density-CMB lensing cross correlation for different redshift bins fit for $b_{\rm gal, 0}$. Table~\ref{tab:dc2_sample} lists the fitted $b_{\rm gal, 0}$ values for each of the variant magnitude limit. For the fiducial magnitude cut, the measured bias is fairly close to the values given by LSST DESC SRD.

\begin{table}
    \centering
    \begin{tabular}{cccccc}
    $\Delta m_{\rm lim}$ & $-1.0$ & $-0.5$    & $0.0$ & $+0.5$ & $+1.0$  \\
    \hline
    \multicolumn{6}{l}{DES} \\
    $b_{\rm gal,0}$&  -- &  1.16 & 1.16 & 1.09 & 1.04 \\ 
    $\Delta_0$ & -- & 0.0297 & 0.03 & 0.0304 & 0.0303 \\
    \hline
    \multicolumn{6}{l}{LSST Y1} \\
    $b_{\rm gal, 0}$& 1.07 & 1.02 & 1.01 & 0.95 & 0.91\\
    $\Delta_0$ & 0.00464 & 0.00473 & 0.005 & 0.00516 & 0.00570\\
    \hline
    \multicolumn{6}{l}{LSST Y10} \\
    $b_{\rm gal,0}$& 1.03 & 1.00 & 0.94 & 0.91 & 0.83\\
    $\Delta_0$ & 0.00251 & 0.00260 & 0.003 & 0.00375 & 0.00456\\
    \end{tabular}
    \caption{The coefficients for galaxy bias and uncertainty in mean redshift for the variant DES, LSST Y1, and LSST Y10 lens samples when the limiting magnitude changes by $\Delta m_{\rm lim}$. The approach for deriving these values is described in Section~\ref{sec:galaxy_bias}, and Section~\ref{sec:bias_mean_z}. For the DES sample at $\Delta m_{\rm lim}=-1.0$, the number of galaxies is too low in \texttt{CosmoDC2} to reliably derive these parameters.}
    \label{tab:dc2_sample}
\end{table}

\subsubsection{Uncertainty in mean redshift}
\label{sec:bias_mean_z}

 It is not straightforward to estimate uncertainties in the mean redshift in \texttt{CosmoDC2} since the values depend on various external factors such as the number of spectra used for redshift calibration, the underlying selection bias in the spectroscopic data, etc. As a result, we will make the assumption that  
 the uncertainty in mean redshift is proportional to $(1+z)$, the same as the fiducial model in Table~\ref{tab:dc2_sample} assumes: 
\begin{equation}
    \sigma_{z_{b}} = \sigma_{z_{b,0}}(1+z),
\end{equation}
and converges to the values in Table~\ref{tab:SRD} when applying the fiducial selection. We use \texttt{CosmoDC2} only to learn how $\sigma_{z_{b,0}}$  scales with different magnitude cuts.

To do this, for a given magnitude selection, we plot the distribution of the difference between the point-estimate photo-$z$ (\texttt{photoz\_mean}) and the true redshift in each tomographic bin. Assuming the distribution is Gaussian, we fit for its standard deviation $\Delta = \sigma(z_{p}-z_{\rm true})$. We then fit the $\Delta$ values for the tomographic bins by a functional form  
\begin{equation}
    \Delta = \Delta_0(1+z_{\rm mean}),
    \label{eq:Delta}
\end{equation}
where $z_{\rm mean}$ is the mean true redshift of each bin. Finally, the uncertainty in mean redshift, $\sigma_{z_b}$, of this magnitude selection will be defined by multiplying $\sigma_{z_b}$ of the fiducial sample by the ratio of $\Delta_{0}$ of this sample and the fiducial sample. Or
\begin{equation}
    \sigma_{z_b} = \sigma^{\rm fid}_{z,b} \frac{\Delta_{0}}{\Delta^{\rm fid}_{0}},
\end{equation}
where $\sigma^{\rm fid}_{z_b}$ denotes the fiducial $\sigma_{z_b}$ values listed in Table~\ref{tab:SRD}, $\Delta_0$ stands for the measured standard deviation between true-$z$ and photo-$z$ (Equation~\ref{eq:Delta}), and $\Delta_0^{\rm fid}$ is $\Delta_0$ for the fiducial limiting magnitude. This model, albeit simple, roughly captures the trend where fainter galaxies are harder to calibrate and therefore have higher $\sigma_{z,b}$ values. Table~\ref{tab:dc2_sample} lists the $\Delta_{0}$ values for each of the variant magnitude limit.

\section{Analysis}
\label{sec:analysis}

Our primary analysis relies on a simulated likelihood analysis to project the cosmological constraints when different galaxy lens samples are used. We describe below how results in previous sections are combined to obtain the key ingredients for the forecast: the covariance matrix and scale cuts (Section~\ref{sec:cov}) and the inference code (Section~\ref{sec:firecrown}). We also define in Section~\ref{sec:firecrown} the Figure of Merit that we will be using in the next section to quantify the change in cosmological constraints when different galaxy samples are used.

\subsection{Covariance matrix and scale cuts}
\label{sec:cov}

We use a simple Gaussian covariance matrix $\mathcal{C}$ for this work \citep{Schneider2002,Crocce2011} and have fixed the covariance at the fiducial cosmology.
\begin{equation}
    \mathcal{C}[C_{X}^{ij} (\ell), C_{X^{\prime}}^{i^{\prime}j^{\prime}}(\ell^{\prime}))]  = \delta_{\ell \ell^{\prime}} \frac{ D_X^{ii^{\prime}}(\ell) D_{X}^{jj^{\prime}}(\ell)+D_{X}^{ij^{\prime}}(\ell) D_{X}^{ji^{\prime}}(\ell)}{(2 \ell + 1) f_{\rm sky}} 
\label{eq:cov_cl}
\end{equation}
where $i,j$ and $i^{\prime},j^{\prime}$ denotes the redshift bin pairs associated with the two considered power spectrum; $X$ is either $\gamma \kappa_{\rm CMB}$ or $\delta_{g} \kappa_{\rm CMB}$. $D_{X}^{ij}(\ell)\equiv C_{X}^{ij}(\ell)+N_X^{ij}(\ell)$ is the sum of the signal $C_{X}^{ij}$ and noise power spectra $N_X^{ij}$, $\delta_{\ell \ell^{\prime}}$ is the Kronecker delta function and $f_{\rm sky}$ is the fractional sky coverage. The noise power spectra are assumed to be zero for cross-correlation.
In practice, the covariance is non-Gaussian and cosmology-dependent, though these should be second-order effects at the level of this analysis since we are mainly interested in relative qualitative changes between different Stage-III and Stage-IV datasets \citep[see also e.g.][]{Barreira2018}. 

To convert Equation~\ref{eq:cov_cl} into a real-space covariance we use the software package \textsc{TJPCov}, which implements the approach in \citep{Singh2021} and \textsc{Skylens}\footnote{\url{https://github.com/sukhdeep2/Skylens_public}}.
\begin{equation}
    \mathcal{C}[\xi_{X}^{ij} (\theta), \xi_{X^{\prime}}^{i^{\prime}j^{\prime}}(\theta^{\prime}))]  =  \mathcal{H} \mathcal{C}[C_{X}^{ij} (\ell), C_{X^{\prime}}^{i^{\prime}j^{\prime}}(\ell^{\prime}))] \mathcal{H}^{T},
\label{eq:cov_xi}
\end{equation}
where $\xi_{X}\equiv \langle X \rangle $ in Equations~\ref{eq:shearkcorr} and~\ref{eq:galkcorr}, and $\mathcal{H}$ is the Hankel transform operator described in \citet{Singh2021}.

When we calculate 2-point functions (for both producing theoretical data vectors and running chains), we use multipole moment $\ell$ up to $6\times10^4$. Specifically, we take $\ell$ as integers between 2 and 49 and concatenate with 200 points between 50 and $6\times10^4$ equally spaced in the logarithmic space. In real space, our data vector contains 20 equally-spaced logarithmic angular bins between $2.5$ and $250$ arcmin. 

For both the theoretical modeling and the covariance matrix, there are modelling uncertainties on small scales, which are commonly dealt with by removing the smallest angular bins from the likelihood analysis. Determining scale cuts is a complicated optimization exercise that involves all probes under consideration and their particular systematics \citep[e.g.][]{Krause:2021}. An optimal determination of scale cuts is beyond the scope of this work. Instead, we fix the scale cuts for all our analyses -- using 10 arcmin for \nk{} and 5 arcmin for \gk{}. These values are mainly motivated by the ongoing work studying cross-correlation between DES Y3 and SPT \citep{y3-5x2method}.  While in reality the optimal scale cuts may considerably affect cosmological constraints if they are significantly different from our assumption, we have checked our analysis using a couple different scale cuts, and find our main conclusions do not change qualitatively (see Appendix~\ref{sec:scale_cuts}.)

\subsection{Cosmological inference and Figure of Merit}
\label{sec:firecrown}

The core of this paper relies on forecasting cosmological constraints from a set of \nkgk{} measurements given data specifications. We use the software package \textsc{Firecrown} to perform the cosmological inference. \textsc{Firecrown} is the LSST DESC cosmology inference code, and its development is ongoing. It relies on the Core Cosmology Library \citep[CCL,][]{Chisari2019} for the theoretical modeling of the data vector (including systematic effects), and the sampling functionalities (including Fisher calculation and a number of sampling methods) packaged in a standalone version of \textsc{CosmoSIS} \citep{Zuntz2015}. We will use mostly use the \textsc{Multinest} sampler \citep{multinest} to derive our constraints via Markov Chain Monte Carlo (MCMC), but for a few cases where we require a large number of chains and we are mainly interested in the relative trends (Section~\ref{sec:maglim} and Appendix~\ref{sec:scale_cuts}), we use the Fisher matrix instead. We show in Appendix~\ref{sec:fisher_mcmc} that using the Fisher matrix forecast does not affect the qualitative trends we are interested compared to a full MCMC chain.

\begin{table}
    \centering
    \begin{tabular}{l|l|l}
         Parameter              & Prior & Fiducial value\\
         \hline
         $\Omega_c$             & $N(0.2664, 0.2)$ &  0.2664\\
         $\Omega_b$             & $N(0.0492, 0.006)$ & 0.0492\\
         $\sigma_8$             & $N(0.831, 0.14)$ & 0.831\\
         $h_0$                  & $N(0.6727, 0.063)$ & 0.6727\\
         $n_s$                  & $N(0.9645, 0.08)$ & 0.9645\\
         $w_0$                  & $N(-1.0, 0.8)$ & -1.0 \\
         $w_a$                  & $N(0.0, 2.0)$ & 0.0 \\
         \hline
         $A_0$                  & $U(-5, 5)$ & 0.0 \\
         $\eta_\mathrm{IA}$     & $U(-5, 5)$ &  0.0 \\
         \hline
         DES & & \\
         $b^i$                  & $U(0.5, 3.0)$ &  $1.3/G(z)$ \\
         Lens $\sigma_{z_b}$ & $N(0.0, 0.03(1+z))$  & 0.0\\
         Source $\sigma_{z_b}$ & $N(0.0, 0.01(1+z))$  & 0.0\\
         $\sigma_{\rm m}$ & $N(0.0, 0.01)$  & 0.0\\
          \hline
         LSST Y1 & & \\
         $b^i$                  & $U(0.5, 3.0)$ &  $1.05/G(z)$ \\
         Lens $\sigma_{z_b}$ & $N(0.0, 0.005(1+z))$  & 0.0\\
         Source $\sigma_{z_b}$ & $N(0.0, 0.002(1+z))$  & 0.0\\
         $\sigma_{\rm m}$ & $N(0.0, 0.013)$ & 0.0 \\
          \hline
         LSST Y10 & & \\
         $b^i$                  & $U(0.5, 3.0)$ &  $0.95/G(z)$ \\
         Lens $\sigma_{z_b}$ & $N(0.0, 0.003(1+z))$  & 0.0\\
         Source $\sigma_{z_b}$ & $N(0.0, 0.001(1+z))$  & 0.0\\
         $\sigma_{\rm m}$ & $N(0.0, 0.003)$ & 0.0 \\
    \end{tabular}
    \caption{Parameter values and priors used for the forecasting in this work. The first section lists the cosmological parameters, the second section lists the nuisance parameters that are the same across all datasets. The last three sections show the baseline nuisance parameters that vary between the three datasets, which are also listed in Table~\ref{tab:SRD}. When we explore variations from the baseline samples in Section~\ref{sec:maglim}, the $b^i$ and lens $\sigma_{z_b}$ values will be different from this table.} 
    \label{tab:parameters}
\end{table}

We forecast constraints primarily assuming a flat $\Lambda$CDM model. But we also show the flat $w$CDM constraints in the baseline case, where for $w$CDM we assume a redshift-dependent dark energy equation of state $w(a)=w_0 + (1-a)w_a$ \citep{Albrecht2006}. We choose to focus on $\Lambda$CDM since the \nkgk{} combination is not very constraining in $w$CDM and using $\Lambda$CDM more clearly illustrates the change in constraining power in the different datasets without being strongly affected by the priors of the parameters. But we have also checked in the $w$CDM case that our main conclusions hold. We use the fiducial parameter values and the priors to be the same as the LSST DESC SRD. The fiducial and priors ranges of the parameters are listed in Table~\ref{tab:parameters}.

\begin{figure*}
    \centering
    \includegraphics[width=0.32\linewidth]{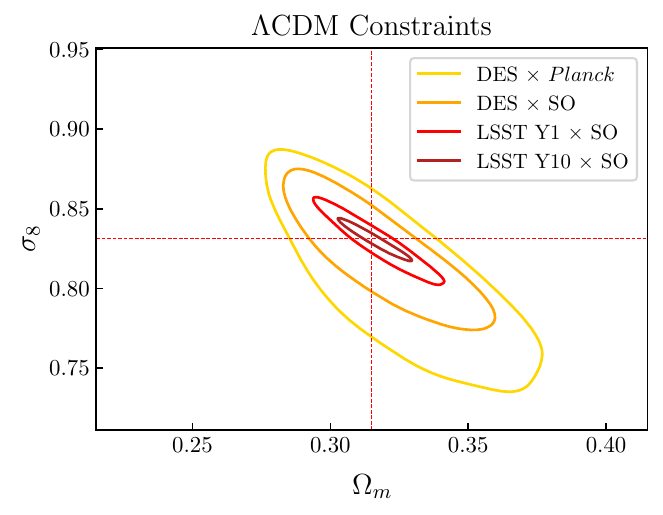}
    \includegraphics[width=0.32\linewidth]{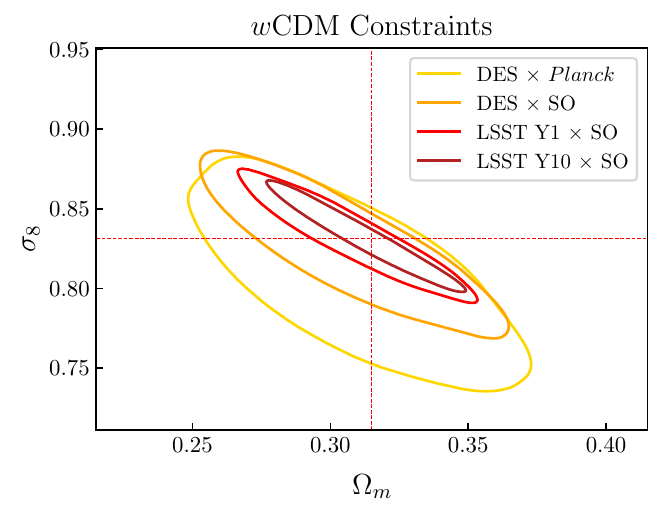}
    \includegraphics[width=0.32\linewidth]{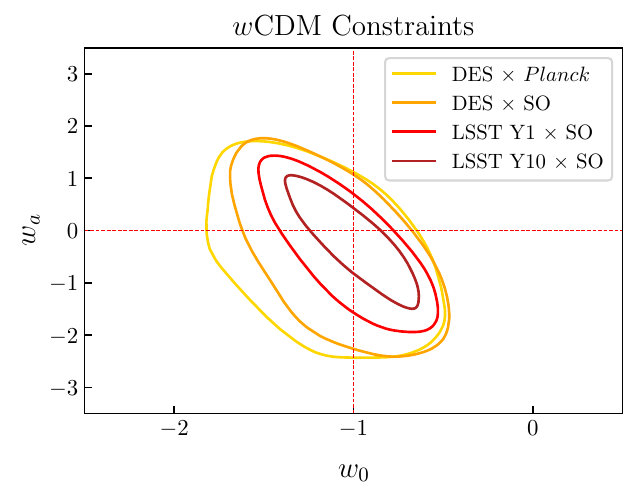}
    \caption{\textit{Left panel}: $\Omega_m-\sigma_8$ constraints from \nkgk{} under $\Lambda$CDM, using the four baseline cases: DES$\times${\it Planck}, DES$\times$SO, LSST Y1$\times$SO, and LSST Y10$\times$SO. \textit{Middle panel}: $\Omega_m-\sigma_8$ constraints with $w$CDM model. \textit{Right panel}: $w_0-w_a$ constraints with $w$CDM model. 
    As we move from DES$\times${\it Planck} to LSST Y10$\times$SO, the constraining power increases significantly for both cosmological models. The contours here show the 1$\sigma$ constraints.}
    \label{fig:baseline_compare}
\end{figure*}

\begin{figure*}
    \centering
    \includegraphics[width=0.45\linewidth]{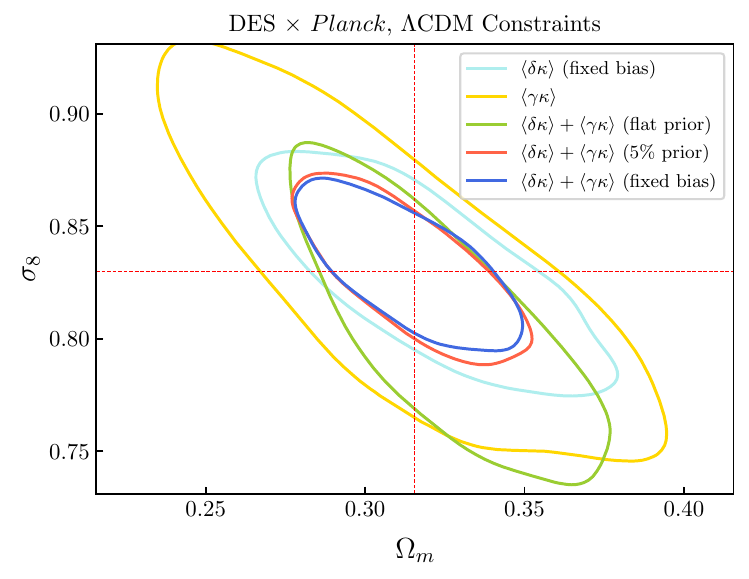}\ \ 
    \includegraphics[width=0.45\linewidth]{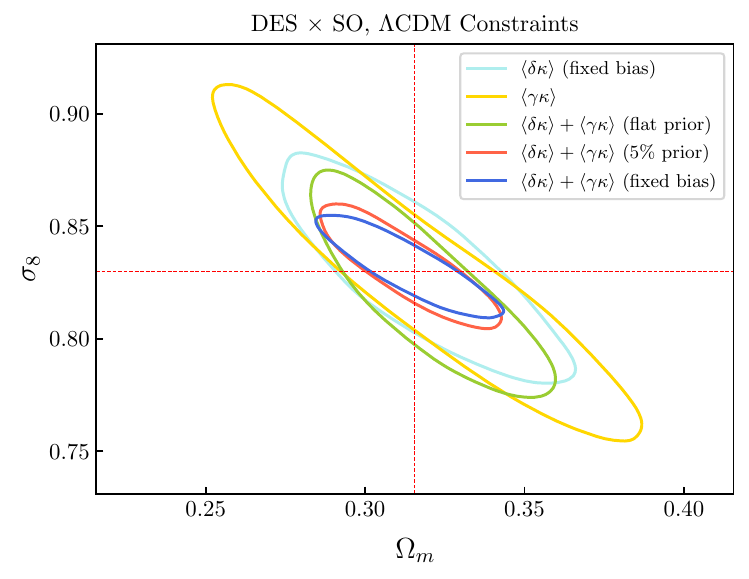}\\
    \includegraphics[width=0.45\linewidth]{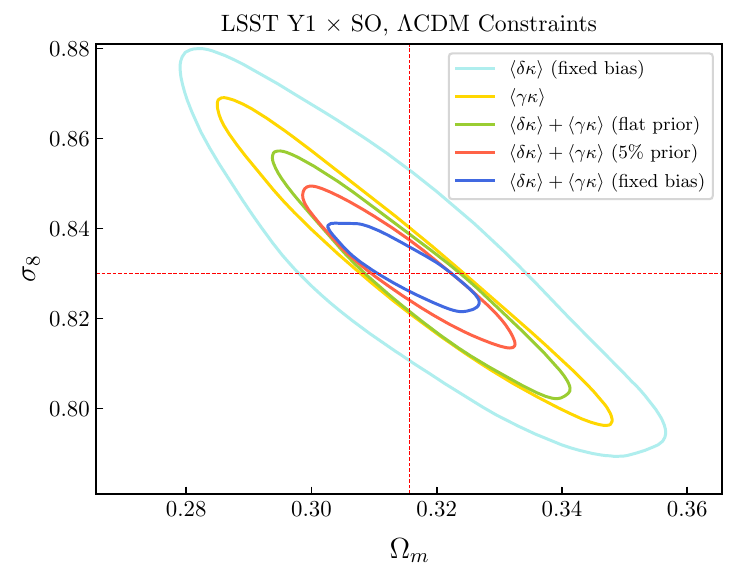}\ \
    \includegraphics[width=0.45\linewidth]{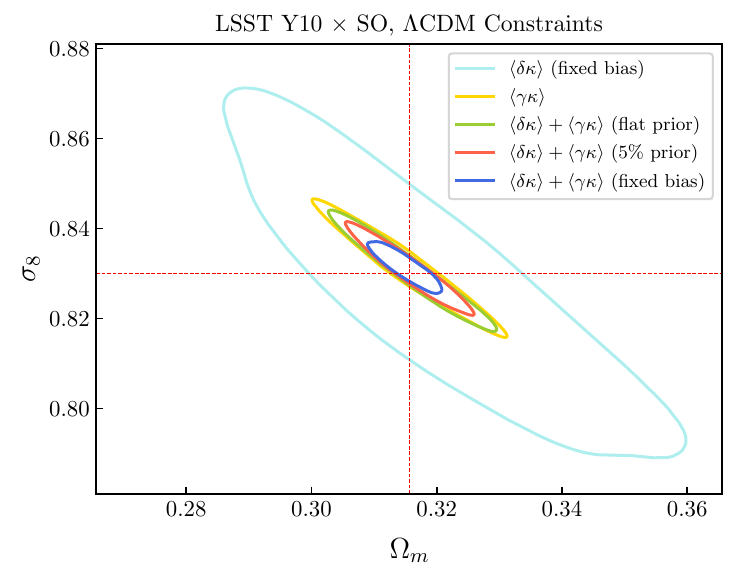}
    \caption{$\Omega_m-\sigma_8$ constraints under $\Lambda$CDM, using the four baseline cases: DES$\times${\it Planck} (top left), DES$\times$SO (top right), LSST Y1$\times$SO (bottom left), and LSST Y10$\times$SO (bottom right). In each panel, there are five curves: \nk{}-only with fixed galaxy bias (cyan), \gk{}-only (yellow), \nkgk{} with flat priors on galaxy bias (green), 5\% prior on galaxy bias (red) and fixed galaxy bias (blue). The legends $\langle\delta\kappa\rangle$ and $\langle\gamma\kappa\rangle$ are shorthands for \nk{} and \gk{} respectively. The contours here show the 1$\sigma$ constraints. We note that the green curves are identical to those in the left panel of Figure~\ref{fig:baseline_compare}.}
    \label{fig:baselines}
\end{figure*}

We quantify the constraining power in each analysis setup by calculating a {\it Figure of Merit} (FoM) defined as:
\begin{equation}
    \mathrm{FoM}=\frac{1}{\sqrt{\det[\mathrm{Cov}(\Omega_m,\sigma_8)]}},
\end{equation}
where Cov refers to the (parameter) covariance matrix between $\Omega_m$ and $\sigma_8$ (the normalization of the matter fluctuations $8h^{-1} Mpc$ scales). This FoM can be intuitively understood as the inverse of the approximate area under the posterior on that parameter sub-space.

\section{Cosmological constraints from galaxy-CMB lensing cross-correlation}
\label{sec:results}

In this section we present the main result of this study -- projection of cosmological constraints from cross-correlation between galaxy density/shear and CMB lensing. First in Section~\ref{sec:fiducial} we show the results for the four dataset combinations introduced in Section~\ref{sec:modeling_samples} and discuss the implication of these cross-correlations moving from Stage-III to Stage-IV galaxy surveys. Next in Section~\ref{sec:srd_cosmodc2} we show how our results change when we use the \texttt{CosmoDC2} mock galaxy catalog to extract more realistic models for the datasets. Using \texttt{CosmoDC2}, we then explore in Section~\ref{sec:maglim} how the constraints change when the lens galaxy sample is varied in terms of the magnitude and redshift selection.

\subsection{Baseline}
\label{sec:fiducial}

For the four dataset combinations (DES$\times$\textit{Planck},  DES$\times$SO, LSST Y1$\times$SO, and LSST Y10$\times$SO), we show in Figure~\ref{fig:baseline_compare} the constraints on the $\Omega_{m}$-$\sigma_8$ plane. These employ the baseline analytic model for each dataset as described in Section~\ref{sec:modeling_samples}. The analysis choices of the cosmological inference follows Section~\ref{sec:analysis}. 

Our main interest in this section is to study how the relative constraining power from \nk{} and \gk{} change between the different data combinations. Figure~\ref{fig:baseline_compare} shows the $\Omega_{m}$-$\sigma_{8}$ constraints for the four cases under $\Lambda$CDM and $w$CDM, as well as the $w_{0}$-$w_{a}$ constraints under $w$CDM. We observe that the constraining power increases significantly between these Stage-III and Stage-IV datasets as expected, and with overall weaker constraints in $w$CDM compared to $\Lambda$CDM. The FoM is 640 (520) for DES$\times$\textit{Planck}, 1220 (740) for DES$\times$SO, 3710 (1370) for LSST Y1$\times$SO, 12300 (1880) for LSST Y10$\times$SO under the $\Lambda$CDM ($w$CDM) model. The gain in constraints under $w$CDM is rather mild going from DES$\times$SO to LSST Y10$\times$SO compared to $\Lambda$CDM, though a similar level of mild increase in constraints can be seen in the $w_0-w_a$ plane (right panel of Figure~\ref{fig:baselines}).

 \begin{figure*}
    \centering
    \includegraphics[width=1.0\linewidth]{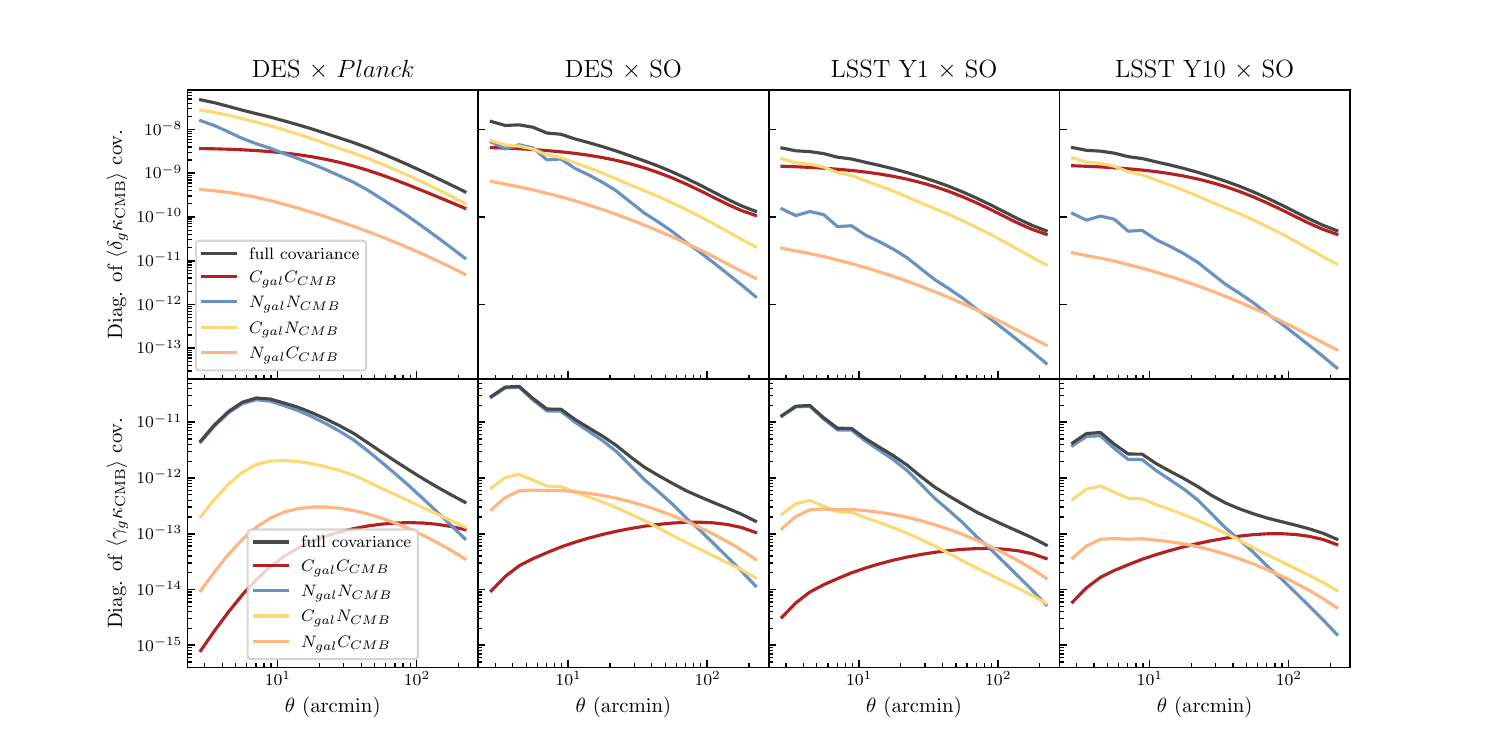}
    \caption{\textit{Top row}: decomposition of the diagonal of \nk{} covariances. \textit{Bottom row}: decomposition of the diagonal of \gk{} covariances. All the covariances are computed using a bin centered around $z=0.7$. In the \nk{} covariance for DES$\times${\it Planck}, the signal-noise term dominates; with an improved CMB noise, the signal-signal term dominates in the large scale, but on small scales, noise-noise, signal-noise, and signal signal terms contribute equally; moving to LSST, the noise-noise term completely fades away in the \nk{} covariance, hence reducing shot noise will not further improve \nk{} constraining power. For the \gk{} covariances, despite that the increasing number density reduces the contribution of the noise-noise term, it still dominates in the small scales. Hence further reducing the shape noise will keep improving \gk{} constraining power.}
    \label{fig:covariance}
\end{figure*}

To further understand the evolution of these constraints, we show in Figure~\ref{fig:baselines} different analysis choices for the $\Lambda$CDM case. In each panel, the ``\nkgk{} (flat prior)'' contours are the same as those in Figure~\ref{fig:baseline_compare}, which we will refer to as the  \textit{fiducial} analysis. We discuss the interpretation of each contour here: 
\begin{itemize}
    \item $\boldsymbol{\langle \delta_g  \kappa_{\rm CMB} \rangle}${\bf -only, fixed galaxy bias:} Since galaxy bias $b_{\rm gal}$ is fully degenerate with $\sigma_8$, freeing $b_{\rm gal}$ significantly decreases the constraining power in the $\Omega_{m}$-$\sigma_{8}$ plane. To get an idea of the constraints coming from $\langle \delta_g \kappa_{\rm CMB} \rangle$ alone we therefore assume fixed galaxy bias. 
    \item $\boldsymbol{\langle \gamma  \kappa_{\rm CMB} \rangle}${\bf-only:} Unlike \nk{}, this combination can constrain $\Omega_{m}$ and $\sigma_{8}$ on its own, though there is a strong degeneracy between these two parameters. 
    \item $\boldsymbol{\langle \delta_g  \kappa_{\rm CMB} \rangle + \langle \gamma  \kappa_{\rm CMB} \rangle}${\bf, flat prior on galaxy bias (fiducial):} This represents a situation where we do not use any external information about the galaxy bias  and purely rely on the constraining power of the \nkgk{} probes alone. Note that this is the fiducial analysis choice, which assumes a flat prior on galaxy bias.
    \item $\boldsymbol{\langle \delta_g  \kappa_{\rm CMB} \rangle + \langle \gamma  \kappa_{\rm CMB} \rangle}${\bf, 5\% Gaussian prior on galaxy bias:} This represents a more optimistic situation, where we have constraints on galaxy bias from additional data. We have made the assumption here that these priors are independent of \nkgk{}, though in some cases the \nkgk{} data vector can be correlated with the dataset that derives the priors. One example is galaxy bias constraints from a \threetwo{} analysis \citep{y3-3x2ptkp}.
    \item $\boldsymbol{\langle \delta_g  \kappa_{\rm CMB} \rangle + \langle \gamma  \kappa_{\rm CMB} \rangle}${\bf, fixed galaxy bias:} This represents the situation where we have already obtained confident constraints of galaxy bias from precedent analysis such as an HOD analysis \citep[e.g.][]{Zacharegkas2021}.    
\end{itemize}

Comparing across the four panels of Figure~\ref{fig:baselines}, we see that relative size and shape between the different contours change. First, comparing the yellow (\gk{}) and the green (\nkgk{}, flat prior) contours, we see changes in constraining power with unchanged degeneracy direction in the $\Omega_{m}$-$\sigma_8$ plane. This improvement comes from the improved constraint in the shear calibration parameters. The addition of \nk{} improves the FoM by 101\% and 94\% for DES$\times${\it Planck} and DES$\times$SO, but that improvement decreases to 57\% once we move to LSST Y1 and further so for LSST Y10, being 25\%. This could also be seen comparing the yellow (\gk{}) and cyan (\nk{}, fixed bias) contours. It is clear that \nk{} is becoming less constraining in comparison with \gk{} when using the galaxy samples from LSST. Noticeably, the cyan contours do not change much going from LSST Y1 to Y10. 

This decrease in the overall contribution of \nk{}  can be explained by looking at the covariance matrix of these different datasets. Figure~\ref{fig:covariance} shows the diagonal of the covariance matrix for a single bin in the \nk{} and \gk{} data vectors separately for the baseline cases. The diagonal covariance matrix is decomposed into four terms as we expand Equation~\ref{eq:cov_cl}: signal-signal ($C_{\rm gal}C_{\rm CMB}$), signal-noise ($C_{\rm gal}N_{\rm CMB}$), noise-signal ($N_{\rm gal}C_{\rm CMB}$), noise-noise ($N_{\rm gal}N_{\rm CMB}$). We observe that going from Stage-III to Stage-IV, the \nk{} covariance is initially dominated by the signal-noise term in the DES$\times${\it Planck} case, but as the CMB lensing map noise decreases, the signal-signal term starts dominating --- the covariance has reached a cosmic variance ``floor''. Meanwhile, since the noise-signal and noise-noise terms are already subdominant in DES $\times$ $Planck$ case, increasing galaxy counts will not help improving the constraining power throughout the four cases. Notice that the limitation due to cosmic variance can be somewhat reduced by using finer redshift bins, but this will also introduce more redshift uncertainty and bias. \nk{} contours from LSST Y1 to LSST Y10 show that doubling the bins can only slightly improve the constraining power (amid the reduced shot noise). We also note that even the DES$\times${\it Planck} combination is very close to being cosmic variance limited, which explains some of the trends we see later in Section~\ref{sec:maglim}. The \gk{} covariance, on the other hand, is noise-noise dominated for all cases. Therefore, it is still possible to increase the constraining power by adding more galaxies to the samples. In summary, the \nk{} and \gk{} trends can be explained by the observation that as we add more galaxies, \nk{} constraint will cease to improve because shot noise is already sub-dominant, but \gk{} constraints are likely to keep improving because of the noise-noise dominance in its covariance.

In addition to the change in the covariance mentioned above, the dominance of \gk{} is also because the \nk{} data vector comes with a free (galaxy bias) parameter per redshift bin. One could argue that for \gk{}, there is the equally unconstrained IA parameters that would weaken its constraints. However, in most analyses the number of free parameters for IA does not change as one increases the number of redshift bins. Interestingly, this would also suggest that moving into Stage-IV experiments, assuming the data characteristics and modeling described in this work, one does not gain much by adding the \nk{} correlation to the data vector -- this could be desirable given the complications of modeling small-scale galaxy clustering signal.

Our conclusion is also supported by the signal-to-noise ratio between the \nk{} and \gk{} data vectors. For the four baseline cases are, i.e. DES $\times$ \planck, DES $\times$ SO, LSST Y1 $\times$ SO, and LSST Y10 $\times$ SO, the \nk{}, \gk{}, and \nkgk{} signal-to-noise ratios are (26, 21, 29), (41, 54, 62), (66, 123, 128), and (73, 231, 233), respectively. We see that initially, the signal-to-noise ratios for \nk{} and \gk{} are comparable -- in fact, as expected from existing work \citep{y3-5x2method,y3-nkgkmeasurement}, \nk{} has slightly higher signal-to-noise ratio than \gk{}. However, as we move into future datasets, the signal-to-noise ratio of \gk{} grows much faster than that of \nk{}. In LSST Y1 and Y10 datasets, we can see that \gk{} has already taken complete dominance in the total signal-to-noise ratio and is much larger than \nk{}.

If we do adopt an IA model with significantly more free parameters \citep[e.g.][or to assume a different IA parameter per redshift bin]{blazek2019}, the \gk{} constraint will become weaker. However, we may also expect that modeling for \nk{} will complicate over time with e.g. more complex galaxy bias models \citep[e.g.][]{goldstein2021perturbation} -- in general, we do not expect LSST Y10 to use the exact same modelling framework as presented here. Thus, it is not straightforward to predict how the relative constraining power change over time beyond the simple approach used here. If the \gk{} constraints become weaker relative to \nk{}, this would delay the time when \gk{} becomes dominant, or even shift the balance between \nk{} and \gk{} altogether. Nevertheless, discussions on cosmic variance above will still hold true.

Comparing the red and the blue ellipses with the green ones in Figure~\ref{fig:baselines} gives us a sense of the additional constraint that we could gain if we have external information on the galaxy bias. We observe two interesting trends: Firstly, the difference between the red and blue contours become larger moving from DES to LSST. This is expected because, at the LSST level, all parameters are more tightly constrained, making the same difference (5\%) between the galaxy bias priors have a much more significant impact on the cosmological constraints. In practice, it is likely that the priors will be tighter when moving from DES to LSST (e.g. if the priors come from \threetwo{}), which will bring the red and blue contours closer. Secondly, it is interesting to see that, although the \nk{} with fixed galaxy bias (cyan) contour becomes much larger than the \gk{} (yellow) contour towards the later stages, there is still nontrivial gain moving from \gk{}-only (yellow) to \nkgk{} fixed galaxy bias (blue). This is somewhat counter-intuitive, but could be explained by the fact that when fixing galaxy bias, \nk{} effectively helps \gk{} constrain both the IA parameters and the shear calibration parameter, which in turn tightens the cosmological constraints.

\subsection{Comparison of baseline SRD and \texttt{CosmoDC2}}
\label{sec:srd_cosmodc2}

Before exploring variations in the baseline samples with \texttt{CosmoDC2}, we compare the characteristics of galaxy samples in SRD with those constructed in \texttt{CosmoDC2} using the same selection criteria. These are not expected to be identical given that the \texttt{CosmoDC2} catalogs underwent a much more extensive validation process with a wider range of datasets \citep{Korytov2019,Kovacs2022}. Furthermore, with \texttt{CosmoDC2} we are able to coherently model all the galaxy properties at the same time -- the catalog naturally contains the correlation between galaxy photometry, redshift, galaxy bias, and nuisance parameters. Finally, the comparison of SRD and \texttt{CosmoDC2} allows us to understand the sensitivity of these forecasting exercises to the detailed assumptions of the galaxy samples.

\begin{table}
    \centering
    \begin{tabular}{l|cccc}
    Dataset & Bin & DES & LSST Y1 & LSST Y10 \\
     & & fid | \texttt{DC2} & fid | \texttt{DC2}& fid | \texttt{DC2}\\
    \hline
    $b_{\rm gal,0}$ & & 1.30 | 1.16 & 1.05 | 0.98 & 0.95 | 0.91 \\
    \hline
    $n_{l}$ (arcmin$^{-2}$) &1 & 0.17 | 0.14 & 2.4 | 2.8 & 2.6 | 1.0 \\ 
                            &2 & 0.24 | 0.10 & 3.2 | 3.3 & 3.5 | 3.2\\
                            &3 & 0.23 | 0.15 & 3.2 | 2.7 & 4.1 | 4.4\\
                            &4 & 0.20 | 0.32 & 3.6 | 4.3 & 4.3 | 3.0\\
                            &5 & 0.15 | 0.25 & 2.1 | 2.3 & 4.3 | 4.3\\
                            &6 & --& --& 4.1 | 5.1 \\
                            &7 & --& --& 3.8 | 6.4\\
                            &8 & --& --& 3.4 | 3.8\\
                            &9 & --& --& 3.0 | 3.0\\
                            &10 &--& --& 2.6 | 4.3\\
    \end{tabular}
    \caption{Comparison of galaxy bias and number density as specified in the baseline lens sample described in Table~\ref{tab:parameters} (first number in each cell of this table, under the columns labeled as ``fid'') and as measured from the \texttt{CosmoDC2} mock galaxy catalog (second number in each cell of this table, under the columns labeled as ``\texttt{DC2}''). }
    \label{tab:cosmodc2_v_srd}
\end{table}

\begin{figure*}
    \centering
    \includegraphics[width=0.24\textwidth]{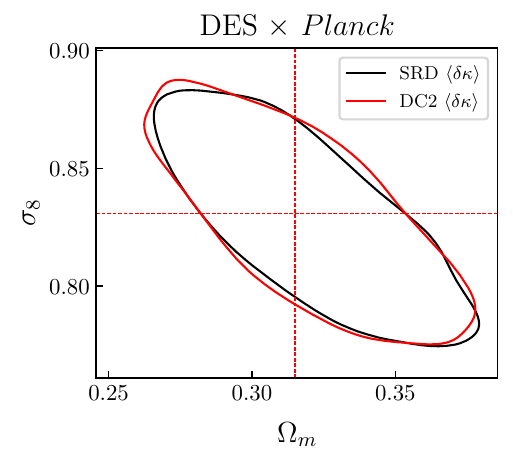}
    \includegraphics[width=0.24\textwidth]{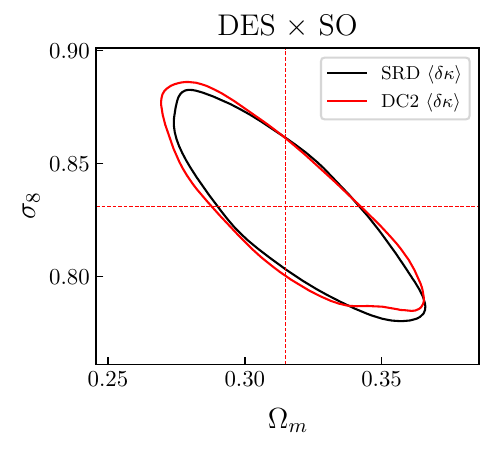}
    \includegraphics[width=0.24\textwidth]{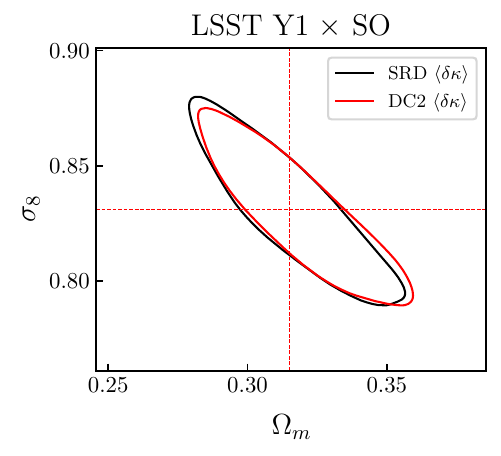}
    \includegraphics[width=0.24\textwidth]{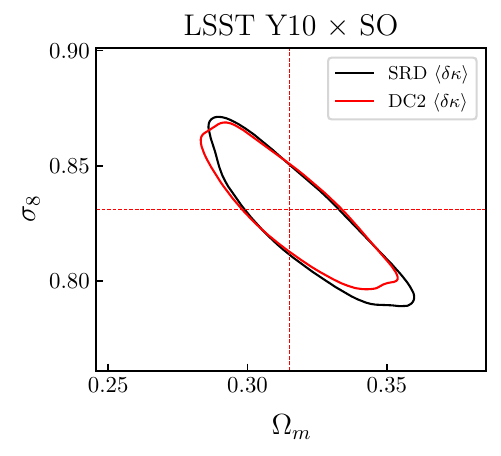}
    \caption{$\Omega_m-\sigma_8$ constraints from \nk{} with $\Lambda$CDM model assuming the galaxy bias is fixed. The red (black) lines show the constraints assuming \texttt{CosmoDC2} (SRD) galaxy samples. We show results for the four baseline cases: DES$\times${\it Planck}, DES$\times$SO, LSST Y1$\times$SO, and LSST Y10$\times$SO.}
    \label{fig:srd_vs_cosmodc2}
\end{figure*}

As already seen in Figure~\ref{fig:Y1_fiducial_nz_cosmodc2}, the $n(z)$ derived from \texttt{CosmoDC2} peaks at higher redshift compared to that of SRD. Given that the SRD model has mainly been checked with the Deep2 catalog, which is a rather small dataset and does not cover the full redshift and magnitude range of our baseline sample, we do not see this level of difference to be surprising. The \texttt{CosmoDC2} $n(z)$'s are also more realistic in the sense that they are less Gaussian and contain outliers far away from the bin center. Next, we compare the lens galaxy bias and number counts in \texttt{CosmoDC2} and the numbers corresponding to Table~\ref{tab:SRD}. Our results are shown in Table~\ref{tab:cosmodc2_v_srd}. We find the values measured from \texttt{CosmoDC2} to be generally lower compared to the SRD values, though the trend between the different datasets is similar. Finally, in Table~\ref{tab:cosmodc2_v_srd} we also list the lens number counts in each bin in both SRD and \texttt{CosmoDC2} -- they on average agree at the 10\% level, though the highest disagreement is at 60\%.  

Taking the baseline galaxy sample from \texttt{CosmoDC2}, we can perform the same forecasting exercise as Section~\ref{sec:fiducial} and check whether our results change qualitatively when using this more realistic sample. Since we are focused on the lens sample, we show in Figure~\ref{fig:srd_vs_cosmodc2} the \nk{} constraints with fixed galaxy bias for the four datasets using the SRD galaxy model and the \texttt{CosmoDC2} galaxy model. We find that despite the difference in the galaxy sample characterization, which includes $n(z)$, number densities, and galaxy bias, the constraints agree extremely well. This serves as a good validation both for the LSST DESC SRD and \texttt{CosmoDC2}: with the more realistic galaxy sample characteristics, we arrive at similar conclusions as the simple analytic model in the SRD; and that the galaxy sample constructed via \texttt{CosmoDC2} agrees with analytical samples in SRD which are based on previous observations.

\subsection{Dependence on magnitude limit and redshift range}
\label{sec:maglim}

We now explore how the constraining power changes when we vary the magnitude limit and redshift range of the lens sample from the baseline. We use the procedure described in Section~\ref{sec:samples} to derive the properties of the variant samples from \texttt{CosmoDC2}. The source sample is kept fix as in Table~\ref{tab:SRD}. To present a clean picture of the trends in the FoM, we use a Fisher forecast here instead of running chains. We notice that \nkgk{} correlation has very loose constraints on IA-parameter $\eta_{\rm IA}$, so we switch to a prior $N(0,5)$ to avoid possible numerical issues in the Fisher calculation.

\begin{figure*}
    \centering
    \includegraphics[width=0.7\linewidth]{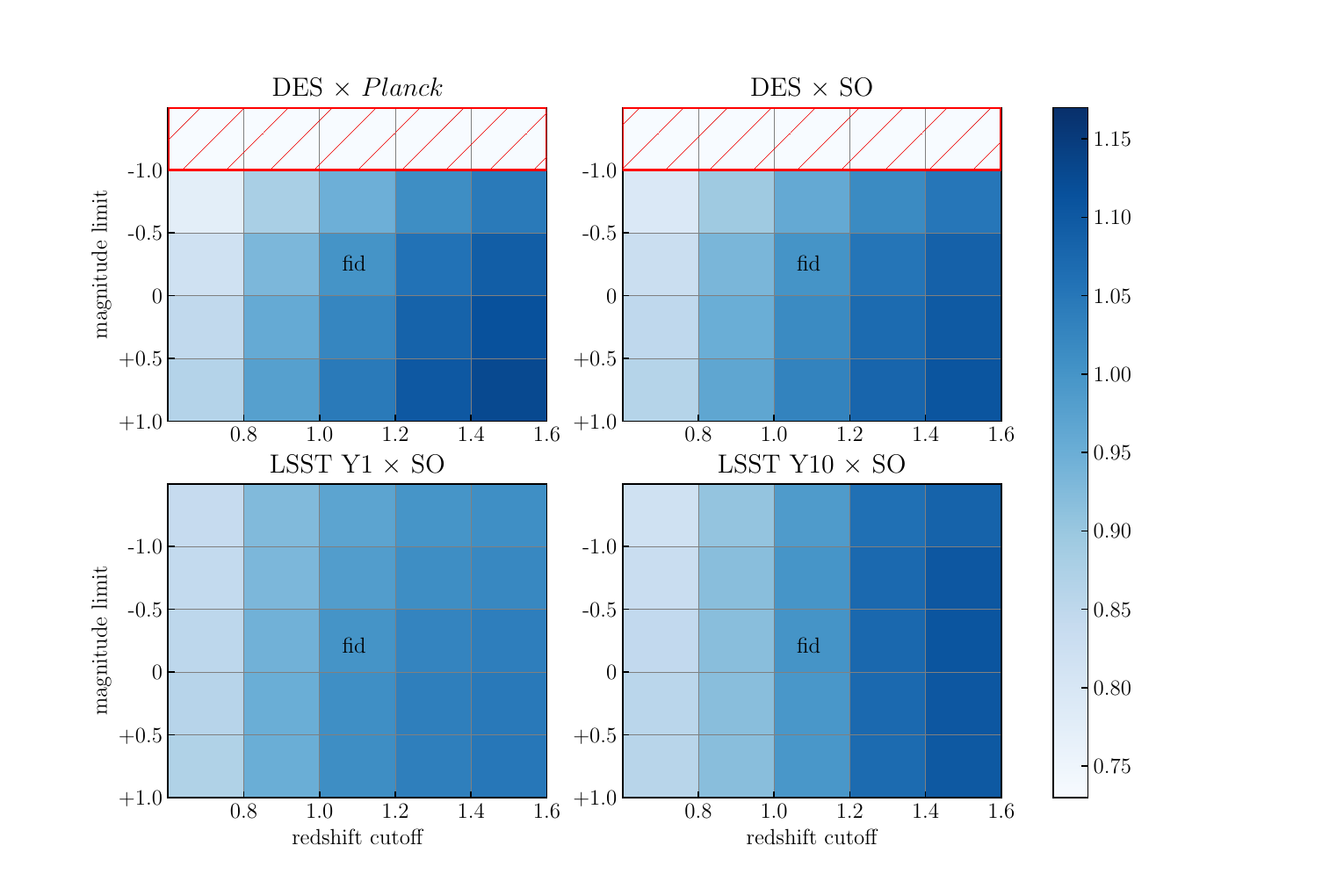}
    \caption{FoM of variant samples for the four baseline cases, normalized against the fiducial samples (labeled by ``fid''). In the DES$\times${\it Planck} case, both redshift range and magnitude limit (and thus number density) have comparable impact on the constraining power. As we moving to LSST Y10, the dependence on magnitude cut (thus number density) fades away. This signifies that shot noise becomes subdominant and the constraining power reaches the cosmic variance limit.}
    \label{fig:variant}
\end{figure*}

Figure~\ref{fig:variant} shows the FoM for the four datasets in a 2D grid when varying both the limiting magnitude and the maximum redshift (and varying the galaxy bias and uncertainty on mean redshifts and their priors according to the models described in Section~\ref{sec:samples}). The FoMs in each panel are normalized by the baseline case (the center cell of the grid), with the relative difference of the baseline in the four panel shown in the left panel of Figure~\ref{fig:baseline_compare}. The colors thus show the relative increase/decrease in the FoM from the fiducial case. The top row of the DES panels are not calculated since the number density of the \texttt{CosmoDC2} catalog at those bright magnitude cuts become too sparse to reliably calculate the nuisance parameters.

\begin{figure}
    \centering
    \includegraphics[width=0.9\linewidth]{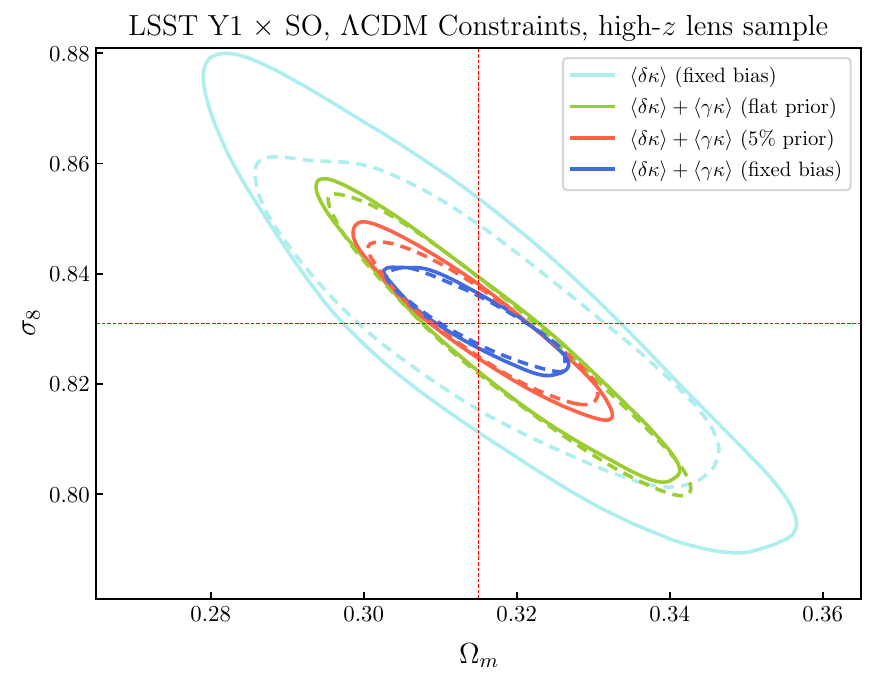}
    \caption{LSST Y1$\times$SO constraints from galaxy-CMB lensing cross-correlation when high-redshift lens galaxies are used (dashed). For reference, the baseline contours are shown in solid lines. The legends $\langle\delta\kappa\rangle$ and $\langle\gamma\kappa\rangle$ are shorthands for \nk{} and \gk{} respectively.}
    \label{fig:highz}
\end{figure}

We observe several interesting trends. First, the general trend of having larger FoM when increasing the number of galaxies (going to fainter limiting magnitudes) and higher redshift is consistent with expectation. Second, the improvements in FoM with magnitude cuts are small, but there tends to be more increase in FoM as we move from brighter to fainter magnitude cuts (moving vertically across the plane) in the cases with DES samples, which can go up to 10\%, than those with LSST samples, with 2\% at most. This is consistent with what we found in Section~\ref{sec:fiducial}, where the constraints provided by the lens samples assumed for DES is close to but not yet cosmic variance-limited, while those of LSST are firmly cosmic variance-limited. Third, the effect of moving to higher redshift fades out from DES lens sample to LSST Y1 lens sample, since \gk{} constraints gradually take dominance. But the effect of adding redshift bins becomes more prominent again in LSST Y10 lens sample. This is somehow expected because the source $n(z)$ extends to higher redshifts (see Figure~\ref{fig:Y1_fiducial_nz}) and the redshift binning is finer. This means that LSST Y10 has more constraining power and more information can be extracted in the redshift direction. But overall, the change in relative FoM is quite small for all cases in the figure, which is roughly contained within $\pm20\%$. This means that the baseline samples assumed in this work is not too far from being ``optimal''. One can improve the \nkgk{} constraints only slightly by making adjustments to the sample. 

Finally, extending from the previous point, we tried to increase the redshift range even further to effectively simulate the cases where special high-redshift samples are constructed \citep[e.g.][]{Krolewski2021} to maximize the overlap with the CMB lensing kernel -- we remove the upper bound of the the highest redshift bin so the $n(z)$ still has nontrivial contributions at $z\sim2$. We would like to understand if, studied under the same framework, these high-redshift samples do indeed constrain cosmology more effectively. In Figure~\ref{fig:highz} we show the case for LSST Y1$\times$SO. We find that the high-redshift lens bins do contribute significantly to \nk{} (compare solid and dashed cyan contours), but when combined with \gk{} the effect is not significant, again due to the fact that in these cases \gk{} dominates the constraints. This also suggests for cases where \gk{} is not yet so dominant (i.e. DES), the addition of the high-z sample could indeed add significant constraining power to the \nkgk{} combination.

\subsection{Combination with other tracers}
While \nkgk{} is worth studying on its own, it is also useful when combined with other tracers e.g. \threetwo{}. For example, as Figure.~\ref{fig:baselines} shows, if the \threetwo{} can strongly constrain bias, \nk{} could in principle become more constraining. We briefly discuss the applicability of our conclusions under this context and directions for future works.

The main conclusion that lens samples should be optimized in terms of nuisance parameters instead of the shot noise is expected to hold. A similar finding has been presented in \citet{y3-2x2maglimforecast} under a different context. In that paper, the authors set out to optimize the lens sample used for  $\langle\delta_g\delta_g\rangle+\langle\delta_g\gamma\rangle$ using DES Y3 data. They found that the lens galaxy samples currently used by DES (similar to the DES baseline sample used in this paper) are close to reaching the cosmic variance limit. Notice that the sample used by that work has limiting magnitude of 22.5. This conclusion will hold with the LSST samples of higher limiting magnitude. Thus, together with \citet{y3-2x2maglimforecast}, our work suggests that going forward in Stage-IV surveys, it is more important to select lens samples with well-understood systematic properties (such as photometric redshift and galaxy bias) than increasing the number counts.

The exact gain in cosmological constraints when adding \nkgk{} to the \threetwo{} probes for future datasets depends heavily on the assumptions in modeling choices and our assumptions on the nuisance parameters. We will therefore leave it for future studies.

\section{Summary}
\label{sec:summary}

In this work, we systematically studied how the constraining power from \nk{} (galaxy density $\times$ CMB lensing convergence) and \gk{} (galaxy weak lensing shear $\times$ CMB lensing convergence) changes as galaxy samples transition from Stage-III to Stage-IV experiments. We investigated both the ``baseline'' cases that assume the same galaxy samples as in 3 $\times$ 2 pt analysis in the LSST DESC Science Requirement Document (SRD), as well as ``variant samples'' with different redshift ranges and magnitude limits. We performed simulated likelihood analyses for each of the samples to forecast cosmological constraining power. The main advances of this work from previous studies are:
\begin{itemize}
    \item For our galaxy sample, we use a realistic mock galaxy catalog \texttt{CosmoDC2} to extract realistic redshift distributions, redshift uncertainties, number densities, galaxy bias and the correlation between them. This approach also serves as a cross-check on the models used in the SRD.
    \item We systematically look at the progression of the cosmological constraints from Stage-III, which we are currently analyzing, to Stage-IV, where we typically make assumptions and extrapolations (for LSST, the values are summarized in the SRD). This allows us to identify factors that could lead to qualitative changes in the cross-correlation constraints going from Stage-III to Stage-IV experiments. 
\end{itemize}

For concreteness, we look at the cosmological constraints for four data combinations: DES$\times${\it Planck}, DES$\times$SO, LSST Y1$\times$SO, and LSST Y10$\times$SO. Below are the main findings of our study: 
\begin{itemize}
    \item Moving from Stage-III (DES$\times$\textit{Planck}) to Stage-IV (LSST Y10$\times$SO), we expect a 20-fold and 3.5-fold increment in \nkgk{} constraining power in the $\Omega_{m}$-$\sigma_{8}$ plane for $\Lambda$CDM and $w$CDM models respectively. Constraints from \nk{} is relatively constant among all baseline cases, while the improvement mostly comes from \gk{}. 
    \item The contributions of \nk{} and \gk{} are comparable at the current stage, but as we approach to LSST Y10$\times$SO, the \nk{} constraint gradually fades out (the improvements of adding \nk{} to \gk{} decreases from 100\% to 25\%).
    \item The \nk{} constraints are cosmic variance limited in Stage-IV (and not far from being cosmic variance limited in Stage-III), so further reducing shot noise in the lens galaxy sample will not gain constraining power. This suggests that for the lens galaxies, one should focus more on improving systematic uncertainties (e.g. prior on galaxy bias or redshift uncertainties). On the other hand, \gk{} constraints do not reach the cosmic variance limit even in LSST Y10; thus further reducing shape noise (increasing number density of source galaxies) can improve constraints. This is a similar strategy suggested for \threetwo{} analyses in previous work.
    \item The forecasted cosmological constraints using the data model from SRD is consistent with that using a data model derived from the  mock galaxy  catalog \texttt{CosmoDC2}. This strengthens the robustness of the main conclusions of this work against detailed assumptions of the galaxy samples.
    \item The above conclusions are applicable to a wide range of samples as we checked the ``variants'' from \texttt{CosmoDC2}. We showed that similar behavior is expected for samples with different redshift ranges (even without an upper bound) and magnitude limits.
    \item We point out that some of the findings above is driven by the sharp increase in the expected lens galaxy number density as well as the much smaller uncertainty in the photometric redshift estimates assumed in the SRD. This perhaps motivates a deeper look at the lens sample specification in the SRD.
\end{itemize}

\noindent We note that we have made several assumptions in this work: 
\begin{itemize}
    \item For the photo-z uncertainty and shear calibration uncertainty, we adopted the ``requirements'' in the SRD instead of estimated values based on methodologies expected for LSST.
    \item We have ignored foreground contamination to the CMB lensing map from e.g. thermal Sunyaev–Zeldovich effect and cosmic infrared background.
    \item We have assumed a simple, scale-independent model for galaxy bias. In reality, the galaxy bias may differ significantly from a linear model in the small scales. This would imply that a constant 10-arcmin scale cut for \nk{} is not entirely realistic. Nevertheless, Appendix~\ref{sec:ultrafaint}~and~\ref{sec:scale_cuts} show that our conclusion is robust against both different biases and scale cuts.
    \item We have assumed a simple 2-parameter intrinsic alignment (IA) model, which could be insufficient in describing the true IA. A more generic model would weaken the constraining power of \gk{} as the IA parameters are fairly unconstrained. This would then change the relative contribution of \nk{} and \gk{} in the combination.
    \item We assumed a 10 arcmin scale cut on \nk{} and 5 arcmin on \gk{}, whereas in reality, determining scale cuts involves specific systematic effects that go beyond the scope of this work. (Though see Appendix~\ref{sec:scale_cuts} for a discussion of our assumption.)
    \item We use a cosmology-independent simple Gaussian covariance matrix in this work. Non-Gaussianity, cosmology-dependence and other complications to the covariance could affect our results slightly.
\end{itemize}

The \nkgk{} cross-correlation combination serves as a powerful consistency test for galaxy-only or CMB-only analyses that is relatively immune to systematic effects. This is particularly valuable given some of the mild tensions seen in current galaxy and CMB experiments. As such, we set out to perform a focused in-depth study on the cross correlations on their own in this work. As we transition from Stage-III to Stage-IV experiments in both galaxy and CMB surveys, we find drastic improvements in the constraining power of \nkgk{} with some unexpected trends. At the same time, we highlight the importance to continuously revisit and realign our forecasts with ongoing analyses in order to achieve the most realistic picture of the future.

\section*{Acknowledgement}

This paper has undergone internal review in the LSST Dark Energy Science Collaboration. 
The internal reviewers were Andrew Hearin, Jonathan Blazek, and Cyrille Doux. 
We thank David Alonso, Matthew Becker, Judit Prat, Mike Jarvis, Javier Sanchez, Suhkdeep Singh, and Ariel Amsellem for help in Namaster, Firecrown and Treecorr. We also thank Yuuki Omori for helpful discussions on CMB lensing models, as well as text-editing and proof reading.
ZZ and CC are supported by DOE grant DE-SC0021949. 
The work of PL at Argonne National Laboratory was supported under the U.S. DOE contract DE-AC02-06CH11357. This research used resources of the Argonne Leadership Computing Facility, which is a DOE Office of Science User Facility supported under Contract DE-AC02-06CH11357.

The DESC acknowledges ongoing support from the Institut National de 
Physique Nucl\'eaire et de Physique des Particules in France; the 
Science \& Technology Facilities Council in the United Kingdom; and the
Department of Energy, the National Science Foundation, and the LSST 
Corporation in the United States.  DESC uses resources of the IN2P3 
Computing Center (CC-IN2P3--Lyon/Villeurbanne - France) funded by the 
Centre National de la Recherche Scientifique; the National Energy 
Research Scientific Computing Center, a DOE Office of Science User 
Facility supported by the Office of Science of the U.S.\ Department of
Energy under Contract No.\ DE-AC02-05CH11231; STFC DiRAC HPC Facilities, 
funded by UK BIS National E-infrastructure capital grants; and the UK 
particle physics grid, supported by the GridPP Collaboration.  This 
work was performed in part under DOE Contract DE-AC02-76SF00515. 

The contributions from the authors are listed below. \\
Zhuoqi (Jackie) Zhang: Carried out most of the analysis and made all the figures. Did significant amount of writing of the text.\\
Chihway Chang: Developed the initial idea of the paper and helped guide the project direction and interpret the analysis results. Did significant amount of writing and polishing of the text. \\
Patricia Larsen: Generated the CMB lensing convergence map in Section~\ref{sec:samples}, helped with early development of the theory code, and contributed to text editing. \\
Lucas Secco: Helped with understanding the behaviour of the covariance matrix as well as text editing.\\
Joe Zuntz: Helped with various technical aspects of \textsc{TXPipe}, \textsc{Firecrown} and \textsc{TJPCov}.

\section*{Data Availability Statement}
No new data were generated or analysed in support of this research.

\bibliography{ref}
\appendix

\section{\texttt{CosmoDC2} Unresolved Faint Galaxies}
\label{sec:ultrafaint}

\begin{figure*}
    \centering
    \includegraphics[width=0.9\textwidth]{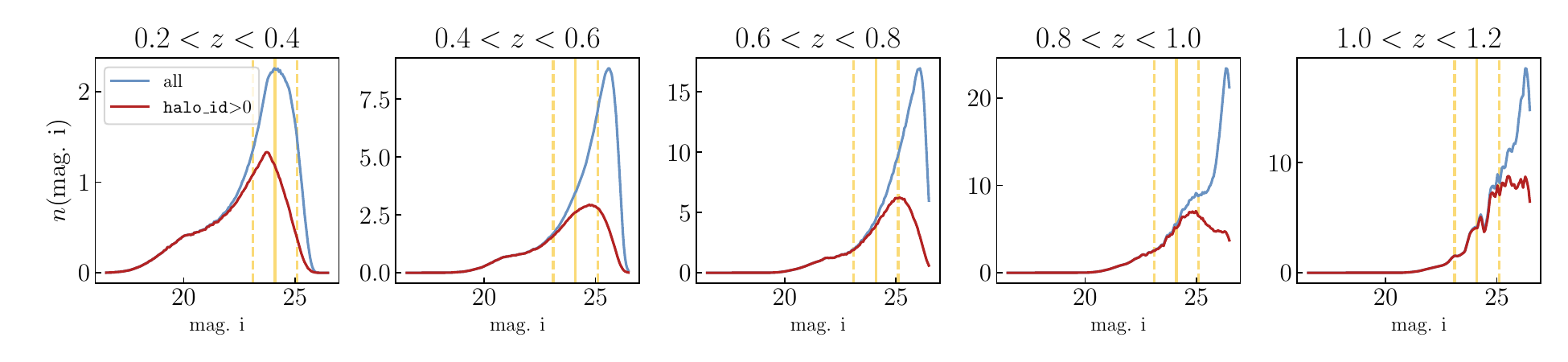}
    \caption{Number of \texttt{CosmoDC2} galaxies that are associated with halos (red, \texttt{halo\_id}$>$0) compared to all the galaxies as a function of magnitude (\texttt{mag\_i}). The five panels show the five baseline lens bins for the LSST Y1 case. The $y$-axis shows the number of galaxies per unit of magnitude in units of one million. The vertical solid lines indicated the baseline magnitude selection, while the two dashed lines show the range of variant samples we consider in terms of limiting magnitude. } 
    \label{fig:haloid0}
\end{figure*}

As discussed in Section~\ref{sec:galaxy_bias}, \texttt{CosmoDC2} contains a population of faint galaxies that do not spatially correlate with dark matter halos, and are therefore excluded when we measure galaxy bias for a given sample. In Figure~\ref{fig:haloid0} we show for the LSST Y1 case, the number of galaxies as a function of magnitude with and without requiring the galaxy to be associated with a halo, for the 5 lens bins. We find that at lower redshift, more galaxies in the sample do not have an associated halo. In the baseline case, only the lowest redshift bin sees a noticeable effect -- about 31\% of the galaxies do not have host halo. When the magnitude cut increases, however, higher redshift bins are also affected. In bin 3, about 15\% of the galaxies do not have host halo when we look at the faintest magnitude selection in our variant samples.

First of all, we note that in general, the lower redshift bins should contribute less to the cosmological constraining power as they do not overlap significantly with the CMB lensing kernel. Second, the next effect of excluding these galaxies that do not cluster is an effectively smaller limiting magnitude and larger galaxy bias. However, according to \citet{Nicola2020}, with a change in limiting magnitude of 0.5 would result in a change in galaxy bias $\sim 3\%$. If we assume the extreme case where the galaxy bias is uniformly overestimated by 3\%, for LSST Y1, we find the FoM changes by 2\%. This level of difference should not affect our main conclusions of the paper. 

\section{Scale cuts}
\label{sec:scale_cuts}

In this work, all cosmological analysis on \nk{} and \gk{} are performed assuming 10 and 5 arcmin scale cuts respectively. However, in reality, scale cuts are determined by balancing the allowed systematic uncertainties on small scales with the systematic errors. In addition, forecasting the actual scale cuts for future data is complicated by the fact that we do expect advances also in the theoretical modeling, which could counteract the need for more conservative cuts as statistical uncertainties shrink. One example is when we look at the scale cuts applied in the DES cosmic shear analysis -- going from \citet{Troxel2017} to \citet{y3-cosmicshear1,y3-cosmicshear2}, although the signal-to-noise increased by a factor of $\sim\sqrt{3}$, the scale cuts remained largely the same due to the added model complexity. 

\begin{figure*}
    \centering
    \includegraphics[width=0.24\textwidth]{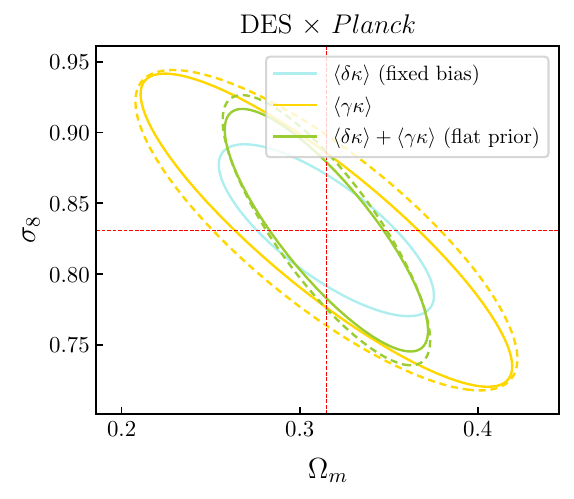}
    \includegraphics[width=0.24\textwidth]{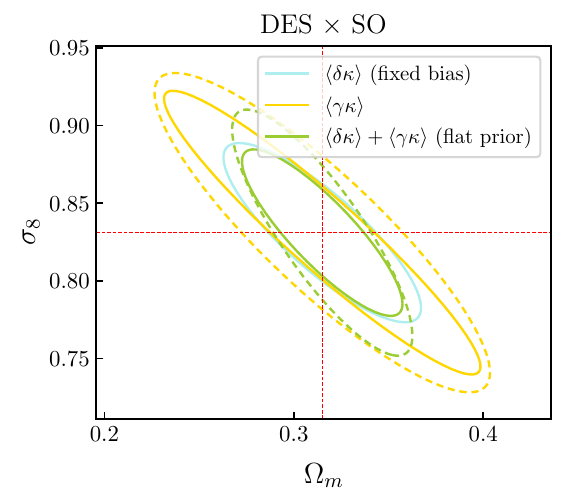}
    \includegraphics[width=0.24\textwidth]{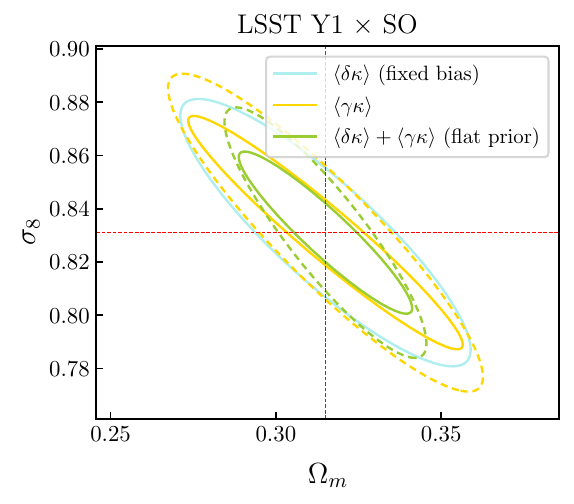}
    \includegraphics[width=0.24\textwidth]{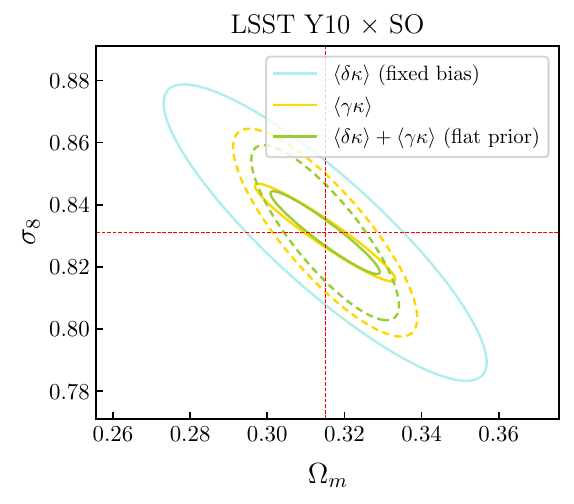}
    \caption{Fisher forecasts for \nk{}, \gk{}, and \nkgk{} (uniform prior). The solid lines are fiducial scale cuts, and the dashed lines are 15 arcmin scale cut for \gk{}.}
    \label{fig:scale_cuts}
\end{figure*}

\begin{figure*}
    \centering
    \includegraphics[width=0.24\textwidth]{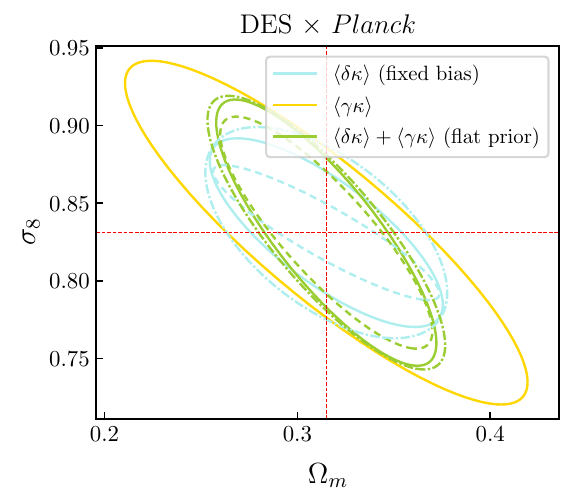}
    \includegraphics[width=0.24\textwidth]{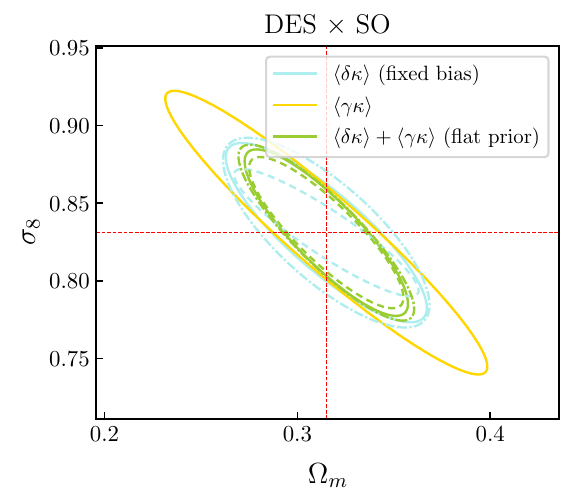}
    \includegraphics[width=0.24\textwidth]{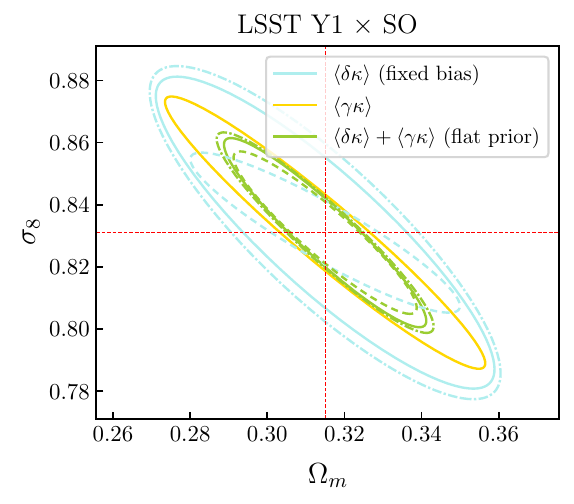}
    \includegraphics[width=0.24\textwidth]{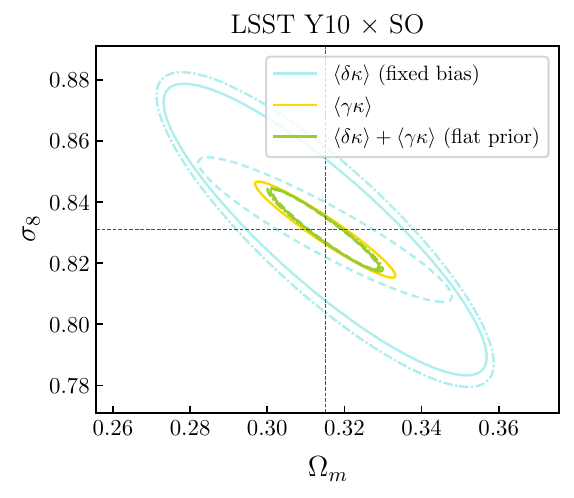}
    \caption{Fisher forecasts for \nk{}, \gk{}, and \nkgk{} (uniform prior) with different scale cuts. The \gk{} scale cut remains at the fiducial 5 arcmin. The \nk{} scale cuts are 5 arcmin (dashed), the fiducial 10 arcmin (solid), and 15 arcmin (dashed-dot).}
    \label{fig:scale_cuts2}
\end{figure*}

We test here how our main conclusions are affected by the particular choice of scale cuts in our baseline analysis. For the the scale cuts on \gk{}, we repeat our fiducial analysis using 15 arcmin instead of 5 arcmin. That is, holding the \nk{} scale cut the same as this work but making the \gk{} scale cut significantly more conservative. The motivation for this extreme scenario is that this weakens the \gk{} constraints, which is dominant in our baseline setting. Figure~\ref{fig:scale_cuts} shows the Fisher forecast constraints with these scale cut. For the scale cuts on \nk{}, we tested with 5 arcmin and 15 arcmin. The motivation is that since our initial estimation of 10 arcmin is roughly an average, the real scale cuts tend to be larger for low redshift bins and smaller for high redshift bins. Fisher forecasts with these scale cuts are shown in Figure~\ref{fig:scale_cuts2}.

We find that, as expected, the constraints strengthen with smaller scale cuts and weaken with larger scale cuts. Nevertheless, in all scenarios, we still see that the \gk{} constraints gradually become more dominant in the \nkgk{} combination as we move from Stage-III to Stage-IV. The main difference between the different scenarios is the exact point  when \gk{} takes dominance. In other words, although we have made a fairly rough choice in the scale cuts for this analysis, we expect all qualitative results in this work to hold, and an error in this choice will mostly shift the point of time at which \gk{} starts to completely dominate the \nkgk{} combination. Even with a very extreme case, we expect by LSST Y10 the contribution from \nk{} to \nkgk{} to be extremely small.

\section{Fisher vs. MCMC Chain}
\label{sec:fisher_mcmc}
\begin{figure}
    \centering
    \includegraphics[width=0.4\textwidth]{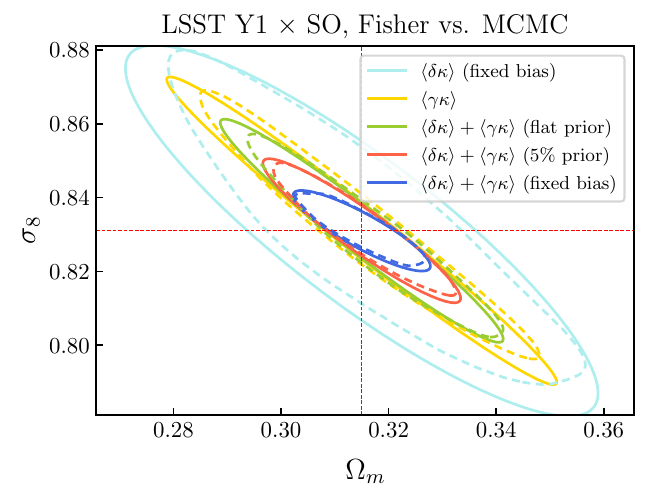}
    \caption{This plot is the same as the bottom-left panel of Figure~\ref{fig:baselines}, but compares Fisher matrix forecast with MCMC sampling. The dashed lines are MCMC chains, while the solid lines are Fisher forecast.}
    \label{fig:Fisher_vs_MCMC}
\end{figure}

In a few occasions in this paper, such as Section~\ref{sec:maglim} and Appendix~\ref{sec:scale_cuts}, we used a Fisher forcast instead of running MCMC chains to estimate cosmological constraints. To validate the prediction from Fisher, we compare the Fisher constraints for the LSST Y1$\times$SO case in Figure~\ref{fig:Fisher_vs_MCMC} with the equivalent MCMC chain constraints in the bottom-left panel of Figure~\ref{fig:baselines}. We can see that despite that the Fisher contours may differ in absolute size from chains, the relative constraints agree with them. Therefore, we believe that Fisher forecasts can reliably reflect the relative trends in constraining power.

\end{document}